\documentclass[12pt]{article} 
 
\usepackage[]{amsmath}
\usepackage[]{graphicx}
\usepackage[]{latexsym}
\usepackage{color}
\usepackage{geometry}
\usepackage{subfigure}
\def\beq{\begin{equation}} 
\def\eeq{\end{equation}} 
\def\barr{\begin{array}} 
\def\earr{\end{array}} 
\def\beqa{\begin{eqnarray*}} 
\def\eeqa{\end{eqnarray*}} 
 
\font\mybb=msbm10 at 12pt 
\def\bb#1{\hbox{\mybb#1}}

\def\b1 {\bb{1}} 
\def\ket#1{| #1 \rangle} 
\def\bra#1{\langle #1 |} 
 
\def\lesssim{\mathrel{\hbox{\rlap{\hbox{\lower4pt\hbox{$\sim$}}}\hbox{$<$}}}} 
\def\gtrsim{\mathrel{\hbox{\rlap{\hbox{\lower4pt\hbox{$\sim$}}}\hbox{$>$}}}}

\makeatletter
\def\equalsfill{$\m@th\mathord=\mkern-7mu
  \cleaders\hbox{$\!\mathord=\!$}\hfill
  \mkern-7mu\mathord=$}
\makeatother

\def\hksqrt{\mathpalette\DHLhksqrt}
\def\DHLhksqrt#1#2{\setbox0=\hbox{$#1\sqrt{#2\,}$}\dimen0=\ht0
  \advance\dimen0-0.2\ht0
  \setbox2=\hbox{\vrule height\ht0 depth -\dimen0}%
{\box0\lower0.4pt\box2}}

\DeclareMathOperator{\erfc}{erfc}
  
\makeatletter
\def\blfootnote{\xdef\@thefnmark{}\@footnotetext}
\makeatother

\begin{document} 

\vspace*{-.6in} \thispagestyle{empty}
\begin{flushright}
\end{flushright}
\vspace{.2in} {\Large
\begin{center}
{\bf Deeply virtual Compton scattering from gauge/gravity duality}
\end{center}}
\vspace{.2in}
\begin{center}
Miguel S. Costa\blfootnote{miguelc@fc.up.pt}, Marko Djuri\'c\blfootnote{djuric@fc.up.pt}
\\
\vspace{.3in} 
\emph{
Centro de F\'\i sica do Porto\\
Departamento de F\'\i sica e Astronomia\\
Faculdade de Ci\^encias da Universidade do Porto\\
Rua do Campo Alegre 687,
4169--007 Porto, Portugal}

\end{center}

\vspace{.3in}

\begin{abstract}
We use gauge/gravity duality to study deeply virtual Compton scattering (DVCS) in the limit of high center of mass  energy
at fixed momentum transfer, corresponding to the limit of low Bjorken $x$, where the process is dominated by the exchange
of the pomeron. Using conformal Regge theory we review the form of the amplitude for pomeron exchange, both at strong and weak 't Hooft coupling.
At strong coupling, the pomeron is described as the graviton Regge trajectory 
in AdS space, with a hard wall to mimic confinement effects. 
This model agrees with HERA data in a large kinematical range.
The behavior of the DVCS cross section for very high energies, inside saturation,
 can be explained by a simple AdS  black disk model. In a restricted kinematical window, this model agrees with HERA data as well. 

\end{abstract}

\newpage

\setcounter{page}{1}

\tableofcontents

\vfill\eject

\section{Introduction}

The approximate conformal symmetry of QCD at high energies gives an ideal ground to test gauge/gravity duality \cite{AdSCFT} in 
situations where there is available data from particle accelerators. Ideally one should be able to reconstruct the dual of QCD
starting from its conformal limit, where the dual geometry is simply AdS space. The fact that the high energy limit of strings in AdS space
does reproduce the observed hard scattering behaviour of QCD \cite{Polchinski:2001tt}, as opposed to that of strings in flat space, 
is a clear indication that  string theory on AdS is indeed a theory of the strong interactions. In this research programme, 
deep inelastic scattering (DIS) in QCD is among the processes that have been extensively studied using the AdS/CFT duality [3-28].  
In particular, there has been progress in reproducing experimental data from HERA in the regime of low Bjorken $x$, where
the gluons dominate the partonic distribution of  the target proton \cite{Saturation,Brower:2010wf}. 
This regime  corresponds to the high energy Regge limit where the interaction between the probe and the target is dominated by pomeron
exchange  \cite{pomeron}. 

\begin{figure}[tbp]
\begin{center}
\includegraphics[scale=0.49]{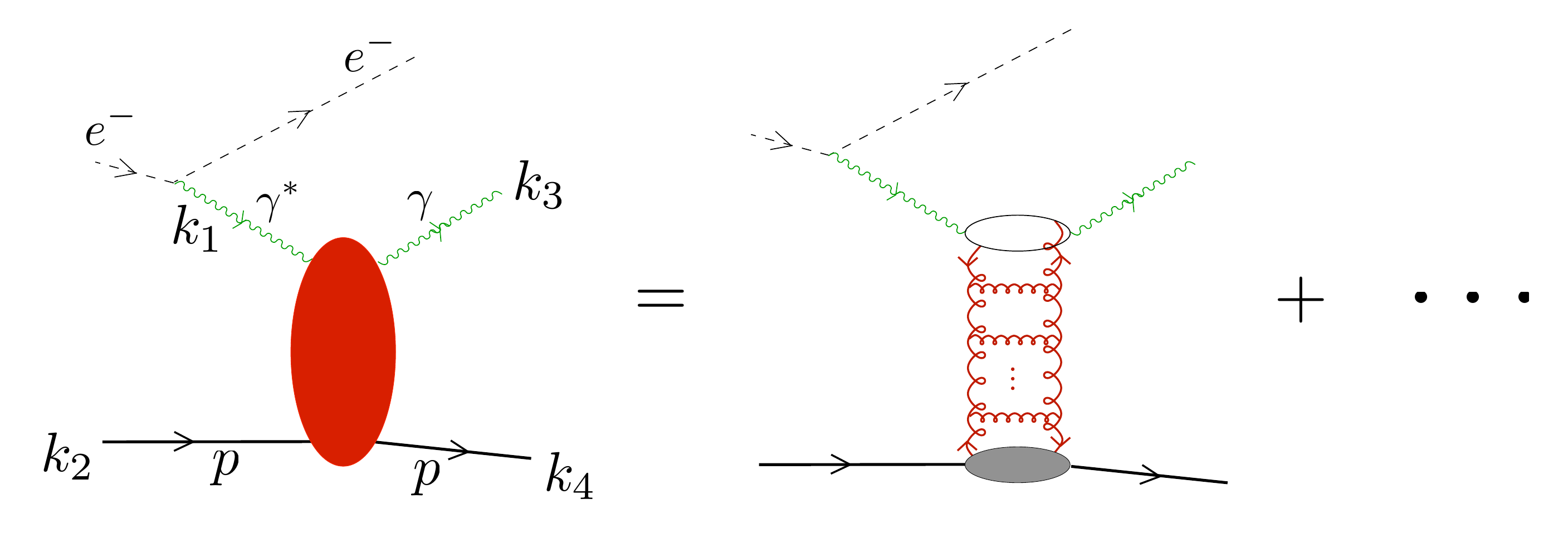}
\caption{{\small In DVCS the electron couples with a basic QED vertex. 
We focus on the QCD process $\gamma^* p \rightarrow \gamma p$, dominated
at weak coupling by the exchange of the BFKL pomeron. There are three kinematical invariants:
the offshellness of the incoming photon $Q^2$; the center of mass energy $W=\sqrt{s}$; 
the momentum transfer $\sqrt{-t}$.}}
\label{fig:DVCS}
\end{center}
\end{figure}

In this paper we shall test gauge/gravity duality models for pomeron exchange in QCD
against the process of deeply virtual Compton scattering (DVCS), where an electron emits a space-like 
off-shell photon that scatters off a scalar target, and is then converted to an on-shell photon. 
Figure \ref{fig:DVCS} represents this basic process. We will be able to match data from ZEUS \cite{Chekanov:2008vy}
and H1 \cite{:2009vda} experiments with very good results. Ultimately we want to make these models a useful tool to match experiments and to
give new predictions for QCD. There have been some works that considered DVCS in gauge/gravity duality \cite{Gao:2009se,Marquet:2010sf,Nishio:2011xz}. 
Here we make direct link to experiment, following our earlier work \cite{Saturation,Brower:2010wf}.

DVCS has a long history of study from QCD. Although a review of this approach is outside the scope of this paper (see for example \cite{Kumericki:2009uq} and references therein, or \cite{Malka:2010zz} for a quick introduction), let us briefly comment on some recent results that relate to pomeron exchange. The authors of \cite{Kowalski:2006hc} use the dipole model to fit the HERA data for exclusive diffractive processes. In this picture the incoming off-shell photon fluctuates into a quark-antiquark dipole, which then interacts with the proton via the exchange of a gluon ladder. They do not use DVCS data to fit their parameters, but still get good agreement when they compare it to their model.  A recent paper \cite{Fazio:2011ex} fits the DVCS (and vector meson production) data using a non-linear pomeron trajectory, with good results. However, we feel an advantage of our approach is that due to our strong coupling starting point we have better access to the non-perturbative region, and due to the diffusion in AdS space of our pomeron kernel we can fit the data for all of the $Q^2$ values, i.e. in both the \em{soft} and the \em{hard} scattering regions, using just a single set of parameters. Furthermore, we fit the differential cross section directly as a function of  kinematic invariants $Q^2, W$ and $t$, and the cross section as a function of $Q^2$ and $W$, without the need to fix one of the variables and only fit the dependence on the other one.  

In section \ref{sec:review} we revisit in some detail gauge/gravity models for pomeron exchange in QCD.
Then in sections \ref{Hadronic} and \ref{sigma}  we do the computations of the cross section  for 
$\gamma^* p \rightarrow \gamma p$, and compare with data in section \ref{Data}. Section \ref{Conc} concludes.

\section{Review of pomeron in gauge/gravity duality}\label{sec:review}

In many $2\rightarrow 2$ QCD processes the Regge behaviour
\begin{equation}
A(s,t) \sim s^{\alpha(t)}\,,\ \ \ \ \ \alpha(t) = \alpha(0) + \frac{\alpha'}{2}\, t \,,
\end{equation}
where $s$ and $t$ are the Mandelstam variables, with $s\gg t$,  is observed.
The exchanged state can be thought of as a reggeon, with the intercept $j_0=\alpha(0) $ determining the growth of total cross sections, $\sigma_{tot}\sim s^{j_0-1}$.
In particular,
this behaviour is observed in  processes  for which a pomeron -- the leading state with the quantum numbers of the vacuum -- is exchanged.
In perturbative  QCD, this story can be made very explicit by considering the hard BFKL pomeron, which is a perturbative resummation in $(\lambda \ln s)^n$, where $\lambda$ is the 't Hooft coupling \cite{pomeron}. 
The hard pomeron describes the exchange of two reggeized gluons, with many ladder diagrams also made of reggeized gluons,
and contains only planar diagrams, so that it is the dominant contribution in a large $N$ expansion (see for instance figure \ref{fig:DVCS}). 
For processes where the QCD scale is important, for instance whenever we probe a target proton at the QCD scale, 
Regge behaviour is still observed, though with a different value of the intercept $j_0$. In this case the  exchanged state is usually called the soft pomeron and it dominates the total
cross section.

Gauge/gravity duality gives a beautiful new insight into the above discussion \cite{Brower1}. 
The main idea is that the hard and soft
pomerons correspond  actually to the very same Regge trajectory, with intercept depending on the effective coupling constant, based on the AdS radial coordinate.
The pomeron Regge trajectory is identified with the BFKL hard pomeron at weak coupling, while at strong coupling it is identified with the  reggeized graviton  trajectory of strings in AdS.  As the coupling is increased,  
the intercept $j_0=j_0(\lambda)$ varies from one to two. 
These ideas can be made very precise in ${\cal N}=4$ SYM, and suggest that one can analyse 
processes dominated by pomeron exchange starting from the conformal limit of QCD \cite{Saturation}. 
In this limit, we have the BFKL hard pomeron as the weak coupling description. 
However, now we also have a strong coupling conformal description, from which we can perturb around. 
More precisely, we may consider a reggeized graviton in $AdS_5$ space and analyse its
predictions for experimental data in regions where strong coupling effects are important \cite{Brower:2010wf}. 

Our starting point is QCD at some fixed value of the coupling 
(in DIS and DVCS, for instance, this is defined by the scale of the probe given by the  off shell photon),
in a conformal kinematical window where all the scales are above the QCD scale.
Let us remark that in this window perturbation theory does not necessarily work, 
since there may be a kinematical enhancement of the effective coupling. This is in fact the case of
low $x$ physics, where non-linear effects give rise to a large gluonic component in the parton distribution functions.

In the conformal limit the momentum space amplitude $A(k_i)$ can be expressed in terms of a reduced conformal amplitude 
${\cal B}(S,L)$, which depends only on two conformal cross ratios $S$ and $L$, that have a natural interpretation
in terms of a dual AdS scattering process. To see how this arises start from the 
four point function of the electromagnetic current operator $j_a(y)$ of dimension $\xi=3$  and of a scalar operator ${\cal O}(y)$ 
of dimension $\Delta$,
\begin{equation}
A(y_i) = \langle j(y_1) {\cal O}(y_2) j(y_3) {\cal O}(y_4) \rangle\,.
\label{AmplitudePosition}
\end{equation}
For simplicity  we suppress, in this section, the indices of the current operator. The full computation is presented in detail in \cite{CCP09}.
The Regge limit corresponds to the limit where the points $y_i$ are sent to null infinity. More concretely, 
defining light-cone coordinates $y=(y^+,y^-, y_\perp)$,
where $y_\perp$ is a point in transverse space $\bb{R}^2$,
this limit is attained by sending
\[
y_1^+\rightarrow -\infty\,,\ \ \ \ \ \ 
y_3^+\rightarrow +\infty\,,\ \ \ \ \ \ 
y_2^-\rightarrow -\infty\,,\ \ \ \ \ \ 
y_4^-\rightarrow +\infty\,,\ \ \ \ \ \ 
\]
while keeping $y_i^2$ and $y_{i\perp}$ fixed. This limit is better studied if we do conformal transformations on the points 
$y_i$, according to
\begin{align}
&x_i= (x_i^+,x_i^-, x_{i\perp})  = - \frac{1}{y_i^+} \left(1, y_i^2, y_{i\perp} \right)\,,\ \ \ \ \ \ \ \ \  i=1,3\,,
\\
&x_i= (x_i^+,x_i^-, x_{i\perp})  = - \frac{1}{y_i^-} \left( 1,y_i^2, y_{i\perp} \right)\,,\ \ \ \ \ \ \ \ \ i=2,4\,.
\end{align}
The conformally transformed amplitude 
\begin{equation}
A(x_i) = \langle j(x_1) {\cal O}(x_2) j(x_3) {\cal O}(x_4) \rangle\,,
\end{equation}
is related to (\ref{AmplitudePosition}) by the standard conformal transformation rules for primary operators. One can then see that in the Regge
limit the amplitude depends only on the small vectors
\begin{equation}
x= x_1 - x_3\,,\ \ \ \ \ \ \ \ \ \ \ \ 
\bar{x}= x_2 - x_4\,.
\label{xxbar}
\end{equation}
Moreover, since conformal symmetry imposes that the amplitude is a function of two cross ratios, 
we can write the only two independent invariants as 
\begin{equation}
\sigma^2 = x^2 \bar{x}^2 \,,\ \ \ \ \ \ \ \ \  \cosh\rho = -\frac{x\cdot\bar{x}}{|x||\bar{x}|} \,,
\label{SigmaRho}
\end{equation}
so that the Regge limit corresponds to sending $\sigma\rightarrow 0$ with $\rho$ fixed.
Finally, we can introduce the Fourier transform  \cite{CCP06}
\begin{align}
A(x,\bar{x})=  \int dp\,  d\bar{p} \,  e^{ -2 i p\cdot x -2 i \bar{p} \cdot \bar{x}}
B(p,\bar{p})\ .
\end{align}
A careful analysis of the $i\epsilon$ prescription for the amplitude $A(x,\bar{x})$ shows
that $B(p,\bar{p})$ only has support  inside the future light cone. We may then 
take $p$ and $\bar{p}$ as future directed timelike vectors, and show that $B(p,\bar{p})$ has the form
\begin{align*}
B(p,\bar{p})  = \frac{\mathcal{B}(S,L) }{
( -p^2 )^{2-\xi}  ( -\bar{p}^2 )^{2-\Delta }     }~.
\end{align*}
The reduced amplitude ${\mathcal{B}(S,L)}$ is extensively discussed bellow, and depends only on the two cross ratios
\begin{align}
S=4|p||\bar{p}| \ ,\ \ \ \ \ \ \ \ \ \ \ \cosh L= -\frac{p\cdot \bar{p}}{|p||\bar{p}|}\ .
\label{SL}
\end{align}
The Regge limit corresponds now to large  $S$ and fixed $L$.

\begin{figure}[tbp]
\begin{center}
\includegraphics[scale=0.3]{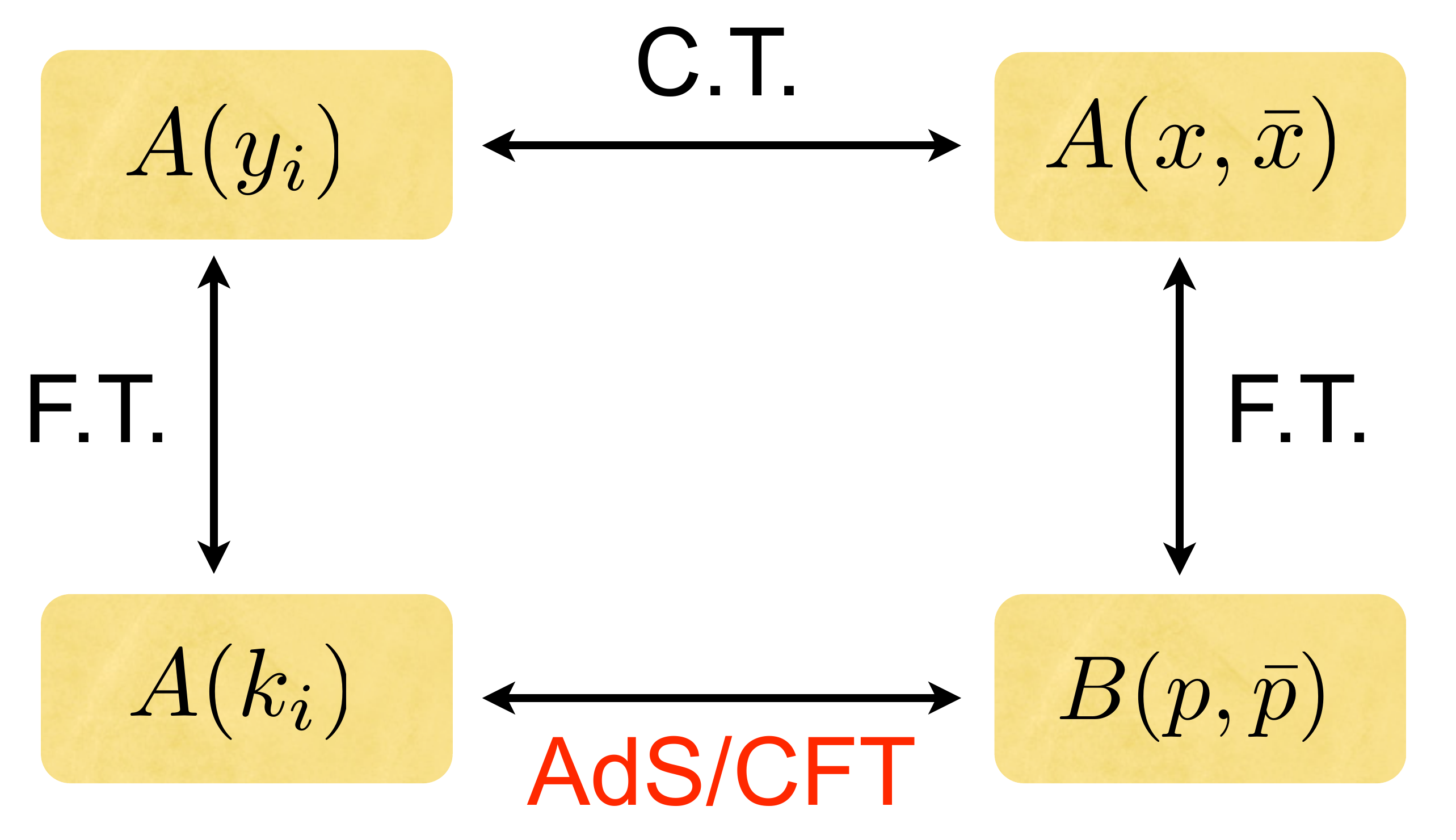}
\caption{{\small As usual the amplitudes $A(k_i)$ and $A(y_i)$ are related by Fourier transform (F.T.). 
In the Regge limit we consider conformal transformations (C.T.) of the points $y_i$, 
defining an amplitude $A(x,\bar{x})$. With a further Fourier transformation we introduce the amplitude $B(p,\bar{p})$.
The relation between the original amplitude $A(k_i)$ and $B(p,\bar{p})$ is the one we would write using AdS/CFT, with the specific function for $B(p,\bar{p})$ depending on the
corresponding Witten diagram.}}
\label{fig:Regge}
\end{center}
\end{figure}

The sequence of transformations, just described, are schematically represented in figure \ref{fig:Regge}. 
The equation relating $A(k_i)$ to ${\cal B}(S,L)$ will only be needed in section \ref{Hadronic}, 
for now we want to discuss the form of the amplitude ${\cal B}(S,L)$ when
it is dominated by the exchange of a reggeon, as dictated by conformal Regge theory \cite{Lorenzo}. 
That  the Regge limit of large $s$ and fixed $t$ corresponds to the kinematic limit of large $S$ and fixed $L$, will also become clear
when we directly relate the cross ratios $S$ and $L$ to the Maldelstam variable $s$ and  the impact parameter $l_\perp$, in section \ref{Hadronic}.
Finally, note that the cross ratios have a natural interpretation from the view point of a dual AdS scattering process, where 
the transverse space is a three-dimensional hyperbolic space $H_3$, whose boundary is conformal to the physical transverse 
space $\bb{R}^2$. The cross ratio $L$ is then  identified with the geodesic distance between two points in $H_3$, and can be related to 
the physical impact parameter  $l_\perp$ along $\bb{R}^2$.  The other cross ratio $S$ measures the local energy squared of the 
scattering process in AdS, which justifies its role as the large energy in the conformal Regge theory reviewed below.

In order to associate a scattering matrix involving a target hadron to the conformal amplitude ${\cal B}(S,L)$, we need to introduce  a discrete spectrum in the theory therefore breaking conformal symmetry. AdS/CFT will instruct us how to do this, without changing the 
conformal structure of ${\cal B}(S,L)$. This will be the case of the AdS black disk and conformal pomeron models considered in this paper. To include the effects of confinement in the amplitude ${\cal B}(S,L)$, we will then consider the 
hard wall pomeron model. For another recent work that explores the conformal symmetry in the context of low $x$
DIS see \cite{Gubser:2011qva}.

\subsection{Regge Theory in CFTs}
\label{CFTRegge}

Let us now see how conformal Regge theory \cite{Lorenzo} restricts the form of the reduced conformal amplitude ${\cal B}(S,L)$ introduced above.
We shall assume that the amplitude for low $x$ DVCS is dominated by a Regge pole, corresponding to the
QCD pomeron. This Regge pole is associated to the exchange of the twist two operators in the leading 
Regge trajectory. The contribution of all these exchanges can then be written as
\begin{equation}
{\cal B}(S,L) =\frac{1}{2S} \sum_{J \ge 2}   \left( S^J + (-S)^J \right) \int d\nu \,  h_J(\nu)\, \Omega_{i\nu}(L)\,.
\label{eq:ReggeTraj}
\end{equation}
This form of the amplitude arises from the  conformal partial wave expansion of the amplitude. In particular,
for fixed spin $J$ of the exchanged operator, the $S$ dependence is given by $S^{J-1}$, and the dependence
in the impact parameter $L$ is determined by the 
coupling of the external operators to the exchanged operator and also by the 
dimension of this exchanged operator, which appears
as a pole of $h_J(\nu)$ at  $i\nu = \Delta(J) - 2 $. The function 
 \begin{equation}
\Omega_{i\nu}(L)= \frac{1}{4\pi^2}\,\frac{\nu \sin({\nu L})}{\sinh L}
\label{Omega}
\end{equation}
is a basis of harmonic functions on a three-dimensional hyperbolic  space $H_3$. For two points on $x,\, \bar{x}\in H_3$, $L$ is defined as the geodesic
distance between these two points. A simple calculation shows that the  function $\Omega_{i\nu}$ satisfies
$(\Box_x + 1+\nu^2) \,\Omega_{i\nu} (x,\bar{x})=0$, and that
\begin{equation}
\int d\nu \,\Omega_{i\nu} (x,\bar{x}) = \delta_{H_3}(x,\bar{x})\,.
\label{H3Delta}
\end{equation}

To compute the sum in (\ref{eq:ReggeTraj}) one re-writes it as a Sommerfeld-Watson integral over spin $J$,
\begin{equation}
{\cal B}(S,L) =- \frac{\pi}{2S}\int \frac{dJ}{2\pi i } \frac{S^J + (-S)^J}{\sin(\pi J)}  \int d\nu \,  h_J(\nu)\, \Omega_{i\nu}(L)\,,
\label{ReggeTraj}
\end{equation}
where the integration contour in $J$ arises from picking all the contributions from the twist two operators of spins $J=2,4,\cdots$.
Deforming the $J$-contour one picks a Regge pole at 
$J=j(\nu)$, which gives the leading behaviour of the amplitude in the Regge limit 
\begin{equation}
{\cal B}(S,L) = \frac{2 \pi i}{N^2} \int d\nu\, \beta(\nu) \, S^{j(\nu)-1}\, \Omega_{i\nu}(L)\,,
\label{Regge}
\end{equation}
Both functions $j(\nu)$ and $\beta(\nu)$ are also functions of the coupling constant. 
The Regge spin $j(\nu)$ is related, by the analytic continuation  $\Delta=2+i\nu$, to the spin--anomalous dimension curve
$\Delta=\Delta(j)$  of the twist two operators on the leading Regge trajectory. 
From the  weak coupling side, both in QCD and in ${\cal N}=4$ SYM, we know that \cite{pomeron}
\begin{equation}
j(\nu,\lambda)  = 1 + \frac{\lambda}{4\pi^2} \left( 2\Psi(1) -\Psi\left(\frac{1+i\nu}{2}\right) - \Psi\left(\frac{1-i\nu}{2}\right)   \right)  + O(\lambda^2)\,,
\label{WeakCoupling}
\end{equation}
where $\lambda=g^2_{YM} N$ is the 't Hooft coupling.
On the other hand, from general arguments, we expect that string theory in $AdS_5$ will give the following strong coupling expansion
for a general class of conformal theories \cite{Brower1}
\begin{equation}
j(\nu,\lambda)  = 2 -  \frac{4+\nu^2}{2\sqrt{\lambda}} + O(1/\lambda)\,,
\label{strong}
\end{equation}
where here the 't Hooft coupling $\lambda=R^4/\alpha'^2$ is defined by the ratio of AdS radius to the dual string $\alpha'$,
and obeys $\lambda\sim g^2_{YM} N$, although the overall numerical factor is only know in specific examples such as ${\cal N}=4$ SYM.

It is straightforward to relate the function $\beta(\nu)$ to the residue of $h_J(\nu)$ in the complex $J$-plane.
The function $\beta(\nu)$   is determined by the reggeon exchange and by the coupling of the reggeon to the external states, 
and can be written in the form
\begin{equation}
\beta(\nu) = W(\nu)\, {\cal G}(\nu) \,\bar{W}(\nu)\,.
\label{beta}
\end{equation}
The functions $W(\nu)$ and $\bar{W}(\nu)$ are impact factors for the external operators  \cite{Cornalba:2008qf},
and were computed at weak coupling for the QCD current operator $j^\mu=\bar{\psi} \gamma^\mu \psi$ 
in \cite{CCP09}. At weak coupling the impact factors can also be directly related to the 
dipole wave functions of the probe and target \cite{Saturation}. 
In ${\cal N}=4$ SYM the next-to-leading order corrections of the impact factors were computed in \cite{Balitsky}.
These functions can also be computed at strong coupling
by assuming minimal coupling between external states and the exchanged string states in AdS \cite{Lorenzo}.
The function ${\cal G}(\nu)$ is related to the exchanged reggeon.

To better understand the Regge residue it is convenient to write the 
amplitude ${\cal B}(S,L) $ in the original form, as a sum over the spin of the exchanged twist two operators, 
\begin{equation}
{\cal B}(S,L) = - \frac{\pi}{2S}\int \frac{dJ}{2\pi i } \frac{S^J + (-S)^J}{\sin(\pi J)} \, {\cal B}(J,L) \,.
\label{MellinJ}
\end{equation}
Equation (\ref{ReggeTraj}) gives the radial Fourier decomposition,
\begin{equation}
{\cal B}(J,L) = \int d\nu\, h_J(\nu)  \, \Omega_{i\nu}(L)\,.
\label{MellinRegge}
\end{equation}
Near the Regge pole $J\sim j(\nu)$ the Fourier component $h_J(\nu)$ will be of the form
\begin{equation}
h_J(\nu)= \frac{2\pi i}{N^2}\, I(\nu)\, G(J, \nu) \,\bar{I}(\nu)\,.
\end{equation}
The impact factors $I(\nu)$ and $\bar{I}(\nu)$ can be simply related to those in (\ref{beta}), after computing 
the $J$ integral. The dependence on $J$ of $G(J, \nu)$ is given  precisely 
by the Regge pole
\begin{equation}
G(J,\nu) = \frac{1}{\nu^2 +  \big( \Delta(J) - 2 \big)^2} \,.
\label{PropJnu}
\end{equation}
For the purpose of computing the $J$ integral by picking the pole at $J=j(\nu)$, we only need to determine the leading behaviour
of this function near the Regge pole, which is defined by the condition $\pm i\nu =  \Delta(j(\nu)) - 2$.
Thus, near the pole we can write
\begin{equation}
G(J,\nu) = \frac{1}{ 2i\nu \Delta'(j(\nu)) \big(J - j(\nu) \big)} \,,
\label{ReggeProp}
\end{equation}
where $\Delta'(J)$ is the derivative of the anomalous dimension function  $\Delta=\Delta(J)$ for the twist two operators in the leading 
Regge trajectory.
At strong coupling this function reduces to the familiar form \cite{Brower1}
\begin{equation}
G(J,\nu) \approx \frac{1}{ 2\sqrt{\lambda} \big(J - j(\nu) \big)} \,,
\end{equation}
but (\ref{ReggeProp}) is valid for any value of the coupling.

The  function $G(J, \nu)$ that gives rise to the Regge pole can be related to the reggeon propagator in AdS space.
To understand this we consider again the radial Fourier decomposition
\begin{equation}
G(J,L) =  \int d\nu\, G(J,\nu)  \, \Omega_{i\nu}(L)\,.
\label{H3DecompositionProp}
\end{equation}
This function is a propagator in the transverse space $H_3$ of the dual AdS scattering process,  satisfying the propagator 
equation
\begin{equation}
\Big[ \Box_{H_3} - 3 -  \Delta(J) \big( \Delta(J) - 4 \big) \Big] G(J,L) =- \delta_{H_3} (x,\bar{x})\,,
\label{H3PropJ}
\end{equation}
where we recall that $L$ is the geodesic distance between $x$ and $\bar{x}$ in $H_3$.

Finally we wish to understand how the above $H_3$ propagator can arrive as the exchange in AdS of a field
of spin $J$ and dimension $\Delta$. When computing the Witten diagram for the exchange of such a field
between two external particles in the Regge limit, one learns that the amplitude depends 
on the integrated propagator of the exchanged field along the null geodesics of each external particles. More
precisely, from the propagator of the exchanged field one obtains
\begin{equation} 
\int_{-\infty}^{\infty} d\lambda d\bar{\lambda} \ \Pi^{(J)}( X(\lambda),\bar{X}(\bar{\lambda})) = -2i S^{J-1} G(J,L)\,,
\end{equation}
where here $\lambda$ and $\bar{\lambda}$ are the affine parameters along the particles geodesics  $X=X(\lambda)$
and $\bar{X}=\bar{X}(\bar{\lambda})$. In this equation the invariant $S$ is given by 
$S=-2 K\cdot \bar{K}$, with  $K$ and $\bar{K}$ the tangent vectors to the 
geodesics.  The propagator $\Pi^{(J)}$ is the contraction of these tangent vectors with the spin $J$  propagator, i.e.
\begin{equation}
\Pi^{(J)} (X,\bar{X}) = (-2)^J K^{\alpha_1} \cdots  K^{\alpha_J}\bar{K}^{\beta_1} \cdots  \bar{K}^{\beta_J} \ \Pi_{\alpha_1\cdots \alpha_J\beta_1\cdots\beta_J} (X,\bar{X}) \,.
\end{equation}
One can then show \cite{Cornalba:2007zb} that equation (\ref{H3PropJ}) for 
the $H_3$ propagator arises from the propagator equation for a
spin $J$ field of  dimension $\Delta$ that, in the Regge limit, is given by 
\begin{equation}
\Big[ \Box_{AdS_5}  -   \Delta(J) \big( \Delta(J) - 4 \big) + J  \Big] \Pi_{\alpha_1\cdots \alpha_J\beta_1\cdots\beta_J} (X,\bar{X}) =
i g_{\alpha_1\left(\beta_1\right.} \cdots  g_{|\alpha_J|\left.\beta_J\right)}  \delta_{AdS_5} (X,\bar{X})+\cdots\,,
\end{equation}
where $\cdots$ represent terms that are negligible.
For instance, in the infinite coupling limit, where only the graviton contributes
to the exchange, by setting  $\Delta(2)=4$  one obtains precisely the propagator that appears in the Regge limit 
of the corresponding Witten diagram for graviton exchange \cite{Cornalba:2007zb,Brower:2007qh}.

\subsection{Pomeron at strong coupling}

Let us finally review the computation of the amplitude ${\cal B}(S,L)$ for pomeron exchange at strong coupling \cite{Brower1,Brower:2007xg}.\footnote{Regge trajectories at strong coupling of other states, for example the odderon, can be similarly found (see \cite{Brower:2008cy} for details).} 
First we compute the $\nu$ integral in (\ref{MellinRegge})
and then the $J$ integral in (\ref{MellinJ}). In particular we are interested in the strong coupling expansion of the reggeon spin
given by (\ref{strong}). Using the explicit expressions for the harmonic functions $\Omega_{i\nu}$ in (\ref{Omega}) and for
the pomeron propagator in (\ref{PropJnu}), 
the $\nu$ integral can be done by residues with the result
\begin{equation}
{\cal B}(J,L) 
= \frac{i }{N^2}\,I(\nu) \,\bar{I}(\nu)\, \frac{e^{- i\nu L} }{2\pi \sinh L}\,,
\label{nuIntegral}
\end{equation}
where in this expression $\nu= \nu(J)= -i (\Delta(J) - 2 )$. To perform the $J$ integral we need to consider first the 
spin-anomalous dimension relation $\Delta=\Delta(J)$, with inverse $J=j(\nu)$ given in (\ref{WeakCoupling}) or (\ref{strong}).
We shall see that the $J$ integral can be done by expanding in $\nu$. Both at weak and strong coupling we can write
\begin{equation}
j(\nu)= j_0 - {\cal D} \nu^2 +\cdots\,,
\label{jExpansion}
\end{equation}
In particular, at strong coupling and keeping only the terms in $1/\sqrt{\lambda}$, the $\nu$ expansion stops precisely at order
$\nu^2$ \cite{Lorenzo}, with
\begin{equation}
j_0=2-\frac{2}{\sqrt{\lambda}}\,,\ \ \ \ \ \ \ \ \ {\cal D}=\frac{1}{2\sqrt{\lambda}}\,.
\end{equation}
Thus, in general the $\nu$ dependence in (\ref{nuIntegral}) is given by
\begin{equation}
 i \nu(J)= \Delta(J)  - 2 = -i \sqrt{  \frac{(j_0-J)}{{\cal D}}} \,.
 \label{nu}
\end{equation}
Hence, to perform the $J$ integral, we note that the integrand (\ref{nuIntegral}) will have a branch cut for ${\rm Re}(J) < j_0$, arising from the argument 
of the exponential. Indeed the impact factors are even functions of $\nu$, admitting an expansion in powers of $\nu^2$, so they do not give rise to any additional 
branch cut. After a simple analysis of the contour of integration we can re-write (\ref{MellinJ}) as 
\begin{equation}
{\cal B}(S,L) =  - \frac{i}{N^2} \,\frac{1}{4\pi \sinh L} \,\frac{1}{S} \int_{-\infty}^{j_0} dJ  \, \frac{S^J + (-S)^J}{\sin(\pi J)} \, \sin (  \nu L)\, I(\nu) \,\bar{I}(\nu)\,.
\end{equation}
Next, using the relation $\nu=\nu(J)$ in (\ref{nu}), we change from the integration variable $J$ to the variable  $\nu$, with the result
\begin{equation}
{\cal B}(S,L) = - \frac{1}{N^2}  \,\frac{{\cal D}\, S^{j_0 -1}}{4\pi \sinh L}  \int_{-\infty}^{+\infty} d\nu \, \nu \, \frac{1 + e^{-i\pi j(\nu)} }{\sin(\pi j(\nu))}\,
e^{- (\ln S)  {\cal D} \nu^2 -i\nu L} \, I(\nu) \,\bar{I}(\nu)\,,
\label{B_integral_nu}
\end{equation}
where we note that $j=j(\nu)$ is given by (\ref{jExpansion}).

We are interested in the high energy limit, at strong coupling, 
with the exchange of all twist two operators in the leading Regge trajectory. This means that we should consider the limit
of ${\cal D} \ln S = \ln S /(2\sqrt{\lambda}) \gg 1$ (this is {\em not} the strict $\lambda \rightarrow \infty$ limit, where only the energy momentum tensor contributes to the exchange).
The previous integral can be done using the saddle point approximation.\footnote{Note that in the weak coupling limit, we must require  ${\cal D}  \ln S \approx  \lambda \ln S  \gg 1$ to compute this integral using the same saddle point approximation.} The saddle point is at
\begin{equation}
\nu_0 = \frac{iL}{2{\cal D}\ln S }\,,
\end{equation}
which is close to zero, so it is consistent with an expansion in  $\nu$.
A straightforward computation gives
\begin{equation}
{\cal B}(S,L) =  -\frac{ i}{\sqrt{\pi}}\,\frac{  I_0 \,\bar{I}_0  }{N^2}\, {\cal D} \left(  \cot\Big(\frac{\pi j_0}{2}\Big) + i \right)  
\,S^{j_0-1}\,  \frac{e^{-\frac{L^2}{4{\cal D}\ln S}}}{  (4{\cal D} \ln S)^{3/2}} \,\frac{ L }{ \sinh L} \,,
\label{Amplitude1}
\end{equation}
where $I_0=I(0)$ is the intercept value for the impact factor $I(\nu)$ and similarly for $\bar{I}(\nu) $. 
In addition to the expansion in $\sqrt{\lambda}/ \ln S \ll 1$ used to compute the $J$ integral, we are also doing 
a large 't Hooft coupling expansion in $1/\sqrt{\lambda} \ll 1$. 

To close this section we consider the overall normalisation  of the amplitude, in particular the dependence on the 't Hooft coupling. 
So far we took $S$ as a dimensionless cross ratio. When we consider a specific conformal field theory with an AdS string dual there 
are two new length scales,  the AdS radius $R$, and the string tension as defined by  $\alpha'$. The AdS radius  allows us to define a dimensionful $S$, 
whose physical interpretation is related to the energy squared with respect to global AdS time. This scale is enough to reproduce the amplitude given
by graviton exchange in AdS from the above formulae, and no $\alpha'$ enters the final result. 
In this case the amplitude can be obtained simply by doing the integral in (\ref{B_integral_nu}) by picking the poles at the zeros
of  $\sin(\pi j(\nu) )\approx -\pi (2-j_0) - \pi {\cal D} \nu^2$. The result is 
\begin{equation}
{\cal B}(S,L) 
=  \frac{i}{N^2}\, I(2i) \,\bar{I}(2i) \, R^2 S\, \frac{e^{- 2 L} }{2 \pi \sinh L}\,.
\label{gravity}
\end{equation}
Here $S$ has dimension of energy squared, as determined by the AdS radius $R$, so we replaced in this expression the original dimensionless
$S$ in  (\ref{B_integral_nu})  by $R^2 S$, since the conformal amplitude ${\cal B}(S,L)$ is dimensionless. 
Looking at the gravity computation done in appendix \ref{SC}, where both for external currents and
scalar fields we have 
\begin{equation}
{\cal B}(S,L) 
= \frac{i \pi}{N^2}\, R^2S\, \frac{e^{- 2 L} }{ 2\sinh L}\,,
\label{gravity2}
\end{equation}
we can read off the impact factors, and in particular obtain the strong coupling result $I(2i)=\bar{I}(2i)=\pi$.

On the other hand, if we consider the graviton Regge trajectory derived from strings in AdS as in (\ref{Amplitude1}),  the string tension plays 
an important role.  Specifically, after introducing the AdS length so that $S$ is dimensionful, 
the dimensionless combination that enters in the above expressions in the place of $S$, 
say at the very beginning in equation (\ref{MellinJ}) for the amplitude,  is precisely  $\alpha' S$. Implementing this modification, and normalizing the overall amplitude
such that it gives the correct gravity result in the limit  $\lambda\rightarrow \infty$, as computed above, equation  (\ref{MellinJ}) is correctly normalized by
\begin{equation}
{\cal B}(S,L) = - \frac{R^2}{\alpha'}\frac{\pi}{2(\alpha'S)}\int \frac{dJ}{2\pi i } \frac{(\alpha'S)^J + (-\alpha'S)^J}{\sin(\pi J)} \, {\cal B}(J,L) \,.
\label{MellinJNew}
\end{equation}
This means that we should replace $S$ by 
\begin{equation}
\alpha' S=\frac{z\bar{z}s}{\sqrt{\lambda}}\,,
\label{alpha'S}
\end{equation}
in the expression (\ref{Amplitude1}), and also multiply this equation by an overall factor $R^2/\alpha'$. Thus,
 our final result for the strong coupling gravi-reggeon amplitude is
\begin{equation}
{\cal B}(S,L) =  \,g_0^2  \left( 1+i \cot\Big(\frac{\pi \rho}{2}\Big) \right)  
\,(\alpha' S)^{1-\rho}\,  \frac{e^{-\frac{L^2}{\rho\ln (\alpha' S)}}}{  (\rho \ln (\alpha' S))^{3/2}} \,\frac{ L }{ \sinh L} \,,
\label{AmplitudeFinal}
\end{equation}
where the parameters $\rho$ and $g_0^2$ are defined by
\begin{equation}
\rho= 2-j_0=4{\cal D}=\frac{2}{\sqrt{\lambda}}\,,\ \ \ \ \ \ \ \ \ \ 
g_0^2 =  \frac{1}{2\sqrt{\pi}} \,\frac{  I_0 \,\bar{I}_0  }{N^2}\,,
\label{g0}
\end{equation}
and we considered  expansions  in $1/ \sqrt{\lambda}\ll1$ and $\sqrt{\lambda}/ \ln S \ll 1$.
The derivation in this section is similar to the one in \cite{Brower:2007xg}, but the way we are presenting the result directly relates the coupling $g_0^2$ to the impact factors
of the external states. 

Let us finally comment on the weak coupling limit of the amplitude, corresponding to the exchange of the hard BFKL pomeron. The very same computation 
leads again to the result (\ref{Amplitude1}), just that this time $j_0 \approx 1 + \lambda \chi(0)$ and ${\cal D} \approx -  \lambda \chi''(0)/2$, where the function
$\chi(\nu)$ can be read off
from (\ref{WeakCoupling}). Taking care of the overall normalization, as in (\ref{MellinJNew}), we obtain
\begin{equation}
{\cal B}(S,L) =  g_0^2 \,2 \sqrt{\lambda}\, {\cal D} \left( 1- \cot\Big(\frac{\pi j_0}{2}\Big)  \right)  
\,(\alpha' S)^{j_0-1}\,  \frac{e^{-\frac{L^2}{4{\cal D}\ln(\alpha'  S)}}}{  (4{\cal D} \ln(\alpha'  S))^{3/2}} \,\frac{ L }{ \sinh L} \,,
\label{AmplitudeFinalWeak}
\end{equation}
where $j_0$ is the weak coupling intercept, $g_0^2$ is still related to the impact factors by (\ref{g0}) and
$(\alpha'  S)$ is given by (\ref{alpha'S}). This expression is obtained by 
considering the  expansions  in  $\lambda\ll 1$ and $1/(\lambda \ln S)\ll 1$.
For instance, we can consider the amplitude between two dipoles. In this case each impact factor comes multiplied by 
$\lambda$, as can be seen from the diagram of figure \ref{fig:DVCS} in the introduction, and as computed explicitly in 
\cite{Cornalba:2008qf,CCP09}. This computation gives therefore the correct leading term at order $\lambda^2$, which is real.

\subsection{AdS black disk from pomeron exchange}
\label{BD}

\begin{figure}[tbp]
\begin{center}
\includegraphics[scale=0.4]{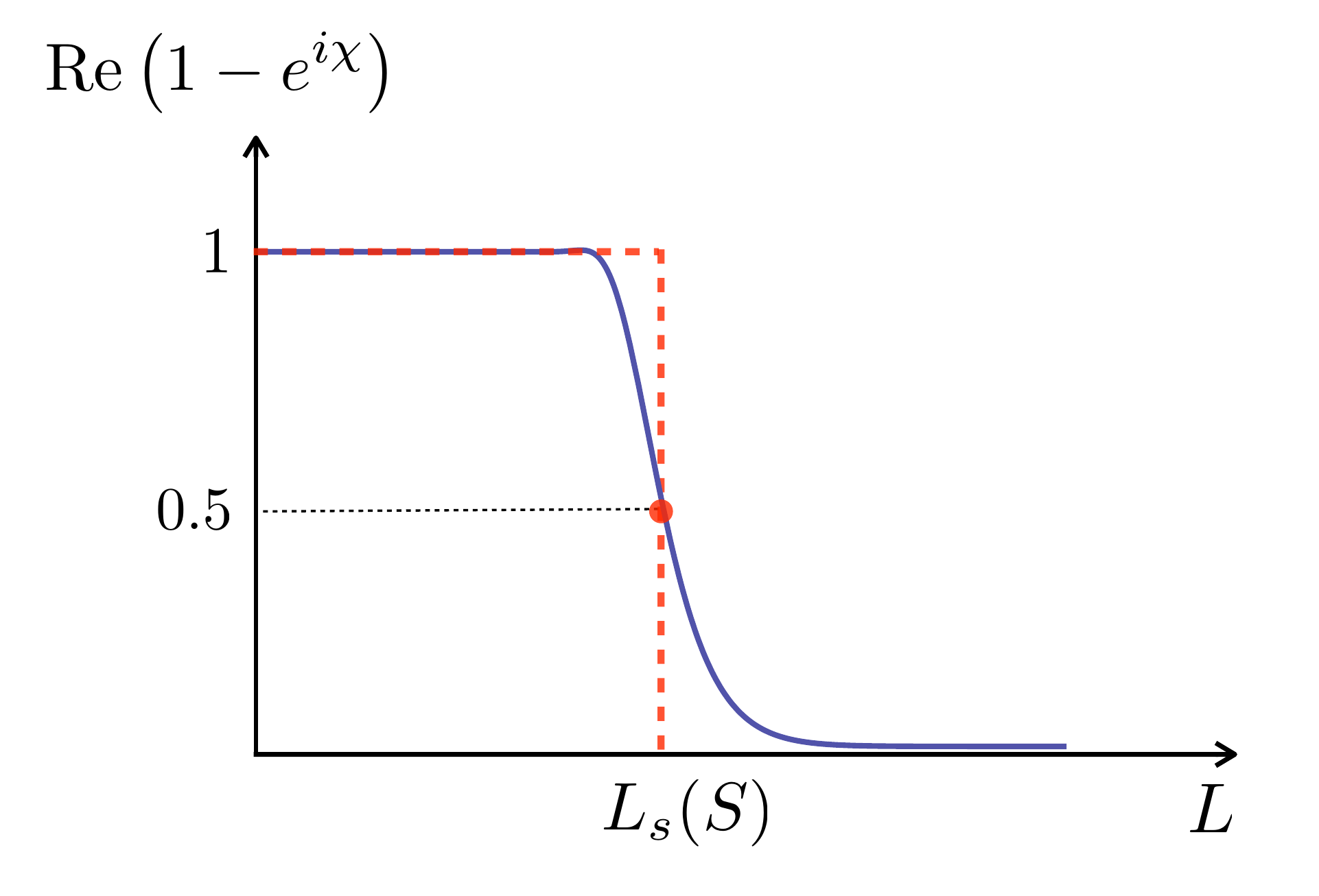}
\caption{{\small Conformal amplitude ${\cal B}(S,L)$ for fixed $S$ obtained from exponentiating the single reggeon exchange.
The picture was obtained from the expression (\ref{AmplitudeFinal}) for the reggeized graviton trajectory.  By varying $S$ it is also simple to verify that the size of the black disk grows  linearly with $\ln S$.}}
\label{BlackDisk}
\end{center}
\end{figure}

Both at weak and strong coupling, one can study the dependence of the amplitude ${\cal B}(S,L)$ with $L$, for fixed $S$. 
In both cases the amplitude reaches unity for some value of $L\equiv L_s(S)$.  One then needs to 
include corrections that come from both multi-ladder reggeon exchange and also from reggeon-reggeon couplings. 
We may then define the amplitude in terms of a phase shift $\chi(S,L)$ such that
\begin{equation}
{\cal B}(S,L) = 1- e^{i\chi(S,L)}\,.
\label{eikonal}
\end{equation}
Provided that the imaginary part of the phase shift grows fast enough as $L$ decreases, the full amplitude 
will in general have the form given in figure \ref{BlackDisk}. This suggests the introduction of a 
conformal (AdS) black disk model for the amplitude, as also plotted in figure \ref{BlackDisk}. The growth of the disk with energy can be determined
by defining first ${\rm Im} \,\chi(S,L_s) \approx 1$. Approximating the full amplitude by the single reggeon exchange (\ref{Regge}),
one can then compute the $\nu$ integral by saddle point, to obtain the general  behaviour $L_s= \omega \ln S$, for
some constant $\omega$. This computation can be done very explicitly
both at weak coupling \cite{Saturation}, for the BFKL pomeron, or at strong coupling, for the BPST pomeron (reggeized AdS graviton). 

For the AdS black disk model we simply assume that the imaginary part of the phase shift grows fast enough 
with decreasing impact parameter $L$, therefore leading to a black disk, independently of the details of the phase shift 
and of the value of the coupling constant.  
For instance,  figure  \ref{BlackDisk} uses for the phase shift the amplitude for the
exchange of a single AdS pomeron (\ref{AmplitudeFinal}). 
Simple combinatorics show that exponentiating the single exchange  amplitude corresponds to the inclusion of all ladder
reggeon exchanges, while dropping reggeon couplings \cite{Cornalba:2007zb,Brower:2007qh,Brower:2007xg}.

\subsection{Conformal symmetry breaking}
\label{ConfBreaking}

This paper considers exclusive photon production in the Regge limit, when a space-like off-shell photon scatters off a scalar target, 
with the final state given by the target and an on-shell photon, as represented in figure \ref{fig:DVCS} in the Introduction.
As usual the Regge limit is defined by taking  large $s=-(k_1+k_2)^2$ 
with fixed momentum transfer $t= - (k_1+ k_3)^2 = -q_\perp^2$ and virtualities $k_i^2$.

As already explained we shall be interested in the kinematical regime where QCD is approximately conformal. 
At weak coupling this corresponds to the limit where the exchange is dominated by the BFKL hard pomeron, 
while at strong coupling it corresponds to the  exchange of the graviton 
Regge trajectory in the dual AdS space.
To be more precise, our starting point is the 
conformal limit of QCD.  We then break conformal symmetry  
using  the gauge/gravity duality by introducing a hard  wall in the dual AdS
space. Thus we start with the high energy Regge limit
\begin{equation}
s \gg -t\, ,Q^2\,,
\end{equation}
where $Q^2 =k_1^2>0$ is the off-shellness of the incoming photon.
As usual the  Bjorken variable is given by
\begin{equation}
x=-\frac{Q^2}{2 k_1\cdot k_2} \approx 
 \frac{Q^2}{s}\,,
\end{equation}
where the last equality is valid in the high energy Regge limit of $x\ll 1$ here considered. 

Conformal invariance  requires that all the above energy invariants are above the QCD scale $\Lambda_{QCD}\sim0.2\ {\rm GeV}$. In particular,
the condition $Q\gg \Lambda_{QCD}$ means that the photon probe is much smaller that the size of the proton, so that corrections
in $\Lambda_{QCD}/Q$ that break conformal invariance are suppressed. 
Most data used in this paper satisfies this condition (some data has $Q\sim 1.5\ {\rm GeV}$, which is still a few times larger than $\Lambda_{QCD}$).
Moreover, we must also satisfy the condition $-t \gg \Lambda^2_{QCD}$ for the momentum transfer, so that corrections in  $\Lambda^2_{QCD}/t$ are also
suppressed. Again this condition is satisfied for most of the data here considered  ($-t$ will range from $0.1$ to $1$ ${\rm GeV}^2$, so that
we will be above  $\Lambda^2_{QCD}\sim0.04\ {\rm GeV}^2$, and only for the lowest value of $-t$ will the approximate conformal symmetry begin to show signs of breaking down). 

Another source for conformal symmetry breaking comes from the masses of quarks, which also introduce new scales. 
To see this note that the coupling between the off-shell photon and the target is determined by the electromagnetic current operator
$j_a= e \sum_f q_f \bar{\psi}_f \gamma_a \psi_f$, where the sum is over quark flavours.
The off-shell photon creates a quark/anti-quark pair which then couples to the target through the exchange of many gluons. 
We therefore need to work in a range of $Q$ such that $2m_f \ll Q$. For these flavours the quarks contribute to the process and 
can be taken as massless. For the other flavours we need to be below the scale defined by the threshold for pair creation, i.e. $Q\ll 2m_f$, so that these
quark loops can be ignored. 
We shall be working in a kinematical region for which
\begin{equation}
1.5 \lesssim Q \lesssim 8.4 \ {\rm GeV}\,.
\label{Qrange}
\end{equation}
Therefore we can safely take the $u$, $d$ and $s$ quarks as massless. For the 
charm quark, with mass  $m_c = 1.27\ {\rm GeV}$, some of the data points will be in the region
where its mass can not be neglected. 
Since the majority of data points have $Q$ above the $2m_q$ threshold we will assume this quark enters the process and is massless
(for the analysis of differential cross section there will be 4 points below the threshold out of 52 , and for the
total cross section 13 points out of 44).  This is clearly an approximation that enters the assumption of
conformal symmetry.\footnote{It also enters the weak coupling computation of the normalization of the currrent 
operator explained in section \ref{Norm}, but this plays no role in the fit to experimental data.}
Finally, for the bottom quark, we can ignore its existence, except
for a single data point at  $Q = 8.4\ {\rm GeV}$, since $m_b = 4.2\ {\rm GeV}$. Thus, we shall 
analyse a kinematical region where the  four quarks  $u$, $d$, $s$ and $c$ are approximately 
massless and the remaining ones can be neglected.

Both the AdS black disk and conformal pomeron models considered in this paper,
assume that conformal symmetry dictates the form of the amplitude.
The very good matching of experimental data obtained for both models clearly points out that we are indeed analysing  a conformal window of QCD.
Conformal symmetry breaking only enters through the assumption that the theory 
contains a discrete spectrum of states, as necessary to define a target hadron.
For the black disk, the final expression is independent of the coupling of the reggeon to the external states, so that
the model  depends only on the parameter $\omega$ introduced in section \ref{BD}, on a scale 
defined by the target hadron, and on a overall constant determined by the normalization of the current operator two-point function.
This model will match data very successfully in a kinematical window where so far there are only a few available experimental points. 
For the conformal pomeron, we need to include the coupling
of the reggeon to external states. This is one of the parameters of this model that multiplies the 
normalization of the current operator two-point function.
Together with the scale of the target hadron and the value of the intercept $j_0$, this model also has three independent parameters. 
We shall see that the conformal pomeron matches data in a quite broad kinematical range.

To  include QCD effects associated to confinement, 
we proceed phenomenologically by considering the AdS
hard wall pomeron model, where the dual AdS space finishes at some fixed scale, defining an additional fitting parameter,
which should be around $\Lambda_{QCD}$. In this phenomenological model one is then able to take into account 
the dependence of the amplitude on new dimensionless ratios constructed from $\Lambda_{QCD}$, 
like $Q/ \Lambda_{QCD}$  and $t/ \Lambda^2_{QCD}$, which break conformal invariance
(as it is usually the case in gauge/gravity models of QCD, this is a  phenomenological model, since 
 we do not have control in terms of a well defined expansion in these small parameters).
 Thus  we expect this model to be more accurate in regions where either the scale set by the
probe, or the momentum transfer, become close to $\Lambda_{QCD}$.
The fit to DVCS data will confirm these expectations.

\section{Hadronic Compton tensor in the Regge limit} \label{Hadronic}

In this section we shall relate the results presented above for the conformal 
amplitude ${\cal B}(S,L)$, given by the conformal Regge theory, to the amplitude 
for the  $2\rightarrow 2$ process in DVCS, as represented in figure \ref{fig:DVCS}. 
This uses the conformal impact parameter representation introduced
in \cite{CCP06} and extensively discussed in \cite{CCP09}. We then compute the hadronic Compton tensor, which 
is the basic object that allows one to compute the  $\gamma^* p \rightarrow \gamma p$ cross section.

Before starting let us note that in the explicit computations here presented  we use 
light-cone coordinates $(+,-,\perp)$, with metric given by $ds^2= -dx^+ dx^- + dx_\perp^2$, 
where $x_\perp \in \bb{R}^2$ is a vector in the impact parameter space. For the incoming particles we take
\begin{equation}
k_1 = \left(\sqrt{s} ,- \frac{Q^2}{\sqrt{s}},0\right)\,,  
\ \ \ \ \ \ \ \ \ \
k_2 = \left(\frac{M^2}{\sqrt{s}},\sqrt{s},0\right)\,,
\label{k1k2}
\end{equation}
where $M$ is the mass of the target, and the incoming off-shell photon is space-like with $k_1^2=Q^2>0$.
For the outgoing particles
\begin{equation}
k_3 =  - \left(\sqrt{s} ,\frac{q_\perp^2 - Q'^2}{\sqrt{s}}, q_\perp\right)\,,   
\ \ \ \ \ \ \ \ \ \
k_4 = -  \left(\frac{M^2+q_\perp^2}{\sqrt{s}},\sqrt{s},-q_\perp\right)\,.
\label{k3k4}
\end{equation}
For the outgoing photon we consider  $k_3^2=Q'^2>0$, and then take the $Q'\rightarrow 0$ on-shell limit. 
In section \ref{TDVCS} we briefly consider lepton pair production for which  the out-going photon is time-like.
Note that the variable $s$, as defined in the above vectors, coincides with the Mandelstam $s$ 
only in the Regge limit here considered. 

\subsection{Conformal impact parameter representation \label{IPR}}

As already discussed, we define the DVCS amplitude starting from a conformal field theory. 
Conformal symmetry breaking will then be implemented following the geometrical intuition 
gained from the AdS interpretation of the scattering amplitude. The basic object we need to 
consider is the momentum space correlation function
\begin{equation}
(2\pi)^4\, \delta\left( \sum k_j \right) i\, T^{ab}(k_j) =  
\left\langle j^a(k_1) {\cal O}(k_2) j^b(k_3) {\cal O}(k_4) \right\rangle\,,
\label{4pf}
\end{equation}
involving the electromagnetic vector 
current $j^a$ and a scalar primary   ${\cal O}$ of dimension $\Delta$.
In the conformal Regge limit \cite{Lorenzo}, 
this correlation function can be written as \cite{CCP06,Saturation,CCP09}
\begin{equation}
T^{ab}(k_j)  \approx  2 i s  \int dl_\perp\,
e^{iq_\perp\cdot l_\perp} 
 \int \frac{dz}{z^3}\, \frac{d\bar{z}}{\bar{z}^3} \,
\Psi^{ab\ \tau}_{\ \ \mu}(z)\, \Phi (\bar{z})  \,
\left[ 1-e^{i\chi (S,L)}  \right]_{\ \tau}^\mu \,,
\label{IPrep}
\end{equation}
where the phase shift $\chi_{\ \tau}^\mu$ is a tensor  that encodes all the dynamical information and, due to conformal symmetry, depends only on 
the variables
\begin{equation}
S=z\bar{z} s \ ,\ \ \ \ \ \ \ \ \ \ 
\cosh L=\frac{z^2+\bar{z}^2+l_\perp^2}{2z\bar{z}}\,.
\label{S,L}
\end{equation}
In section  \ref{CFTRegge}  we refered to this  conformal amplitude as
${\cal B}(S,L) = 1- e^{i\chi (S,L)}$, and for simplicity we omitted the indices in that section.
 For a conserved current the Greek indices $\mu$ and $\tau$ 
label tangent directions to a three-dimensional hyperbolic space $H_3$, with metric in Poincar\'e coordinates given by
\begin{equation}
ds^2(H_3)=\frac{dz^2+ds^2(\bb{R}^2)}{z^2}\ .
\end{equation}
We work in units such that the AdS radius $R=1$, otherwise there would be an overall 
factor of $R^2$ in this metric, and  the cross ratio $S$ would be given instead by
$S=z\bar{z} s /R^2$.
The scalar function $\Phi$ and the tensor function $\Psi^{ab\ \tau}_{\ \ \mu}$
are associated with the operators ${\cal O}$ and $j^a$. Their explicit form 
was given in \cite{CCP09}.

The above conformal representation (\ref{IPrep}) is quite natural from the view point of the dual AdS scattering process, where
transverse space is precisely a three-dimensional hyperbolic space $H_3$, whose boundary is conformal to the physical transverse 
space $\bb{R}^2$. The cross ratio $L$ is then  identified with the geodesic distance between two points in $H_3$ 
that are separated by $l_\perp$ 
along $\bb{R}^2$ and have radial coordinates $z$ and $\bar{z}$.
The other cross ratio $S$ measures the local energy squared of the scattering process in AdS.
The directions tangent to $H_3$, which we label with 
Greek indices, are now the  physical polarizations of the AdS gauge field dual to the vector  operator $j^a$. 
Moreover the functions $\Psi^{ab\ \tau}_{\ \ \mu}(z)$ and $\Phi (\bar{z})$ can be related to the 
 bulk to boundary propagators (non-normalizable modes produced by a plane wave
source created by the dual operator at the boundary) 
of the dual AdS fields and their couplings to the exchanged reggeon, as shown in the strong coupling
computation (i.e. AdS Witten diagram) presented in appendix \ref{SC}. 

It is important to note that the above  conformal  representation of the amplitude is valid for any value of the coupling constant, since it relies
only on conformal invariance. At weak coupling we can reproduce this exact form of the correlation function, with the phase shift determined by the 
hard pomeron exchange \cite{CCP09}. The functions  $\Psi^{ab\ \tau}_{\ \ \mu}(z)$ and $\Phi (\bar{z})$
can then be related to the dipole wave functions of the current and scalar operators \cite{Saturation}.
At strong coupling this form of the amplitude is reproduced by computing the AdS Witten diagram
given by t-channel graviton exchange. This is a very non-trivial check of the all coupling form of the conformal impact representation (\ref{IPrep}). To verify this fact, and to convince the sceptical reader, we present this computation in appendix \ref{SC}.\footnote{The very same Witten diagram was computed in \cite{Bartels}.}

In the Regge limit the amplitude is dominated by the pomeron Regge
trajectory. At leading order in $\mathcal{N}=4$ SYM, as well as in the $\lambda\rightarrow \infty$ limit in $\mathcal{N}=4$ SYM, the phase shift  $\chi_{\ \tau}^\mu$ is diagonal.
For simplicity, we assume that this holds for any value of the coupling and write, throughout this paper, 
\begin{equation}
\chi_{\ \tau}^\mu =  \chi (S,L)\,\delta_{\ \tau}^\mu\,.
\label{PhaseShift}
\end{equation}
At strong coupling the phase shift is diagonal because the coupling between the gauge field and the graviton in AdS is determined by the energy-momentum tensor, 
which singles out the $T^{++}$ component in the Regge limit. Recalling the  construction of the pomeron vertex operator \cite{Brower1} from the  graviton Regge trajectory in AdS, we 
also expect the amplitude for this Regge trajectory to remain diagonal. 

Before we show how to compute the hadronic tensor that results from the impact parameter representation, let us note that
for a diagonal phase shift the contraction $\Psi^{ab}\equiv\Psi^{ab\ \mu}_{\ \ \mu}(z)$
appearing in (\ref{IPrep}) has the explicit form
\begin{align}
&\Psi^{ab}(z) 
= 
\left(
\begin{array}{ccc}
 s\Psi_0&   \left( Q'^{2} +  q_\perp^2 \right)\Psi_0 & \sqrt{s}\,q_\perp^j   \Psi_0 \\ \\
  Q^{2} \Psi_0&   \frac{1}{s} \, Q^{2}  \left( Q'^{2} +  q_\perp^2 \right)\Psi_0 &    \frac{1}{\sqrt{s}} \,   Q^{2} \,q_\perp^j \Psi_0\\ \\
  0 &\frac{2}{\sqrt{s}}   \, Q Q'     \,q_\perp^i \Psi_1&Q Q'   \,\delta^{ij} \Psi_1
 \end{array}
\right)\,,
\label{Psi}
\end{align}
where
\begin{equation}
\Psi_n=  - C\, \frac{    \pi^{2}  }{6  } z^4 K_{n}(Qz)K_{n}(Q'z)\,,\ \ \ \ \ \ \ \ \ \ \ \ \ \ \ \ (n=0,1)
\label{Psi_n}
\end{equation}
with $K_n$ the modified Bessel function.
For completeness let us also write the explicit form of the scalar function appearing in (\ref{IPrep}),
\begin{equation}
 \Phi (\bar{z}) = \bar{C}\, \frac{2\pi^2}{\Gamma(\Delta)\Gamma(\Delta-1)}  \left(\frac{QQ'}{4}\right)^{\Delta-2} \bar{z}^4
 K_{2-\Delta}(Q\bar{z})
 K_{2-\Delta}(Q'\bar{z})
 \,.
 \label{Phi}
 \end{equation}
Although the reader may recognise in this expression the form of bulk to boundary AdS propagators, let us emphasise again  that these expressions were obtained by
simply assuming conformal symmetry, and are valid for any value of the coupling constant. Finally, note that
the constants $C$  and $\bar{C}$ are fixed by the normalization of the two-point functions for the external operators that give the disconnected 
piece of the four point function (\ref{4pf}), with
\begin{equation}
\langle j^a(y)j^b(0) \rangle= C \,\frac{y^2\eta^{ab}-2y^ay^b}{(y^2+i\epsilon)^4} \,,\ \ \ \ \ \
\langle {\cal O}(y){\cal O}(0) \rangle= \frac{\bar{C} }{(y^2+i\epsilon)^{\Delta}} \ .
\label{2pfNormalization}
\end{equation}

\subsection{Hadronic tensor \label{HT}}

The hadronic Compton tensor is closely related to the above correlation function. It is given by the matrix element
of the time-ordered product of two electromagnetic currents
\begin{equation}
W^{ab}(k_j) =  i \int d^4y\,e^{i k_1 \cdot y}\bra{k_4}{\rm T} \left\{j^a(y)   j^b(0)\right\}  \ket{k_2}\,,
 \label{DefHadronicTensor}
\end{equation}
where $\ket{k_i}$ represents the hadronic state with momentum $k_i$ ($i=2,4$). 

We can now use the impact parameter representation  (\ref{IPrep}) to write the hadronic Compton tensor as 
\begin{equation}
W^{ab}(k_j)  \approx  2 i s  \int dl_\perp\,
 e^{iq_\perp\cdot l_\perp} 
 \int \frac{dz}{z^3}\, \frac{d\bar{z}}{\bar{z}^3} \,
\Psi^{ab}(z)\, \Phi (\bar{z})  \,
\left[ 1-e^{i\chi (S,L)}  \right] \,,
 \label{HadronicTensor}
\end{equation}
where we already assumed a diagonal phase shift as explained in (\ref{PhaseShift}). The subtlety in this equation, 
in comparison with (\ref{IPrep}) for the correlation function (\ref{4pf}), is in the function $\Phi(\bar{z})$ associated to the target.
At strong coupling, as can be seen in the correlation function computation presented in appendix \ref{SC}, this function is related to the product of two bulk to boundary propagators
of the scalar field $\phi$. The gauge/gravity duality gives then a clear geometric picture on how to break conformal symmetry, 
therefore introducing a discrete spectrum in the theory. This is done by deforming the $AdS$ space in the IR region of large holographic variable $z$, and the simplest way to do this is to introduce a hard wall in the AdS space at $z=z_0$ \cite{Polchinski:2001tt}. 
Moreover, just as for strong coupling, the hard wall model can also be defined using the conformal impact parameter representation (\ref{IPrep}), 
valid for any value of the coupling. In particular, this may also define a way of breaking conformal symmetry for the weak coupling BFKL pomeron.
We comment on this point in the Conclusion.

We therefore introduce a IR cut off $z_0$, and follow our strong coupling AdS intuition. 
In this respect there are three aspects that need to be taken into account. 
Firstly, the scalar field $\phi$ has a discrete spectrum with  normalizable modes $\phi_n(\bar{z})$ satisfying 
\begin{equation}
\int \frac{d\bar{z}}{\bar{z}^3}\,  \phi^\star_m(\bar{z}) \phi_n(\bar{z}) = \delta_{m,n}\,.
\end{equation}
The function $\Phi(\bar{z})$ should therefore be determined by one such 
normalizable mode $\phi_n(\bar{z})$. In this paper we do not take into account the details of the target function, instead we shall use the simplest possible function, i.e. we consider the Dirac delta function
\begin{equation}
\Phi(\bar{z}) = \bar{z}^3 \delta( \bar{z} - z_*)\,,
\label{target}
\end{equation}
where $z_*$ defines the scale of the target.
This approximation will simplify considerably our formulas, still allowing for an excellent fit to experimental data.

Secondly, we need to be careful with the IR region when 
$Q'$ is sent to zero, since in this case the bulk to boundary propagator for the field dual to the outgoing on-shell photon
 spreads over the whole AdS space, therefore probing the IR region.
In the particular case of DVCS, for which the outgoing photon is on-shell, we shall see in the next section that
the  $Q'\rightarrow 0$ limit of the amplitude is well defined because we only need to consider transverse incoming and outgoing 
polarizations, so we do not need to be that careful in this respect.
More generally, however, we must impose Neumann boundary conditions at the IR wall, which means that the Bessel
functions defining (\ref{Psi_n}) should be replaced according to \cite{Hong:2004sa}
\begin{align}
K_0(Q' z)& \rightarrow 
K_0(Q' z) -\frac{K_0(Q'z_0)}{I_0(Q' z_0)}\,I_0(Q' z)\,,
\label{ProbeHardWall}
\\
K_1(Q' z) &\rightarrow 
K_1(Q' z) +\frac{K_0(Q'z_0)}{I_0(Q' z_0)}\,I_1(Q' z)\,.
\nonumber
\end{align}
The amplitude defined in this form is well defined, even when $Q'$ becomes small.
The same procedure should also be followed for the incoming off-shell photon, particularly
when $Q$ approaches the QCD scale (this will only happen for a reduced number of data points in section \ref{Data}).

Thirdly, we also need to include the effects of the hard wall in the pomeron propagator \cite{Brower1}, 
as briefly described in section \ref{HWP}.
 
\subsubsection{Normalization\label{Norm}}

We need to correctly normalize the function $\Psi^{ab}$ in the impact parameter representation for the hadronic tensor in 
(\ref{HadronicTensor}), just as the impact parameter representation for the correlation functions is normalised according to
(\ref{Psi}), (\ref{Psi_n}) and (\ref{2pfNormalization}). Again, this can be done be equating the disconnected pieces of (\ref{DefHadronicTensor}) and (\ref{HadronicTensor}).
For (\ref{DefHadronicTensor}) we have
\begin{equation}
W_{disc}^{ab} (k_j) = \langle k_4 | k_2 \rangle\, i C \int d^4 y \,e^{i k_1 \cdot y} \,\frac{y^2\eta^{ab}-2y^ay^b}{(y^2+i\epsilon)^4} \,,
\end{equation}
where we note that $\langle k_4 | k_2 \rangle = (2\pi)^3 \sqrt{s} \,\delta^{(3)} ( {\bf k}_2 + {\bf k}_4)$. 
To recover  this expression from  (\ref{HadronicTensor}) one needs to follow the computation presented in an appendix of \cite{CCP09}.
This computation basically involves the integral representation of the
bulk to boundary propagator derived directly from the CFT Regge theory, without making any reference to AdS space. The final result is exactly as in (\ref{Psi}) and (\ref{Psi_n})
for the four point correlation function, with the normalization of the current two point function given in (\ref{2pfNormalization}). For details we refer the reader  to reference \cite{CCP09}.

It remains to fix the normalization constant $C$. The electromagnetic current operator has the well known form
\begin{equation}
j_a = e \sum_f q_f \bar{\psi}_f \gamma_a \psi_f\,,
\end{equation}
where the sum is over quark flavours, $e$ is  the electron charge and $q_f$ the corresponding quark charge in units of $e$. 
We shall be working in the kinematical region (\ref{Qrange}) already discussed in section \ref{ConfBreaking}.
At weak coupling we can compute the two point function of this current, taking into account only the quark flavours that 
couple the photon probe to the target. Following the discussion in section \ref{ConfBreaking}, in the kinematical range
considered in this paper only the $u$, $d$, $s$ and $c$ quarks contribute. With this assumptions, the constant $C$ computed using
free Wick contractions in QCD is
\begin{equation}
C = e^2\frac{3}{4\pi^4}\, \frac{10}{9} = 7.845 \times 10^{-4}\,. 
\label{C}
\end{equation}

The above is a weak coupling computation. When fitting to experimental data, we are actually in a region where the coupling is not small. More specifically, we will obtain
for the value of the intercept $j_0\sim 1.2 - 1.3$, which clearly is in a region of finite coupling. Moreover, both the conformal pomeron and hard wall pomeron models consider an 
expansion around strong coupling. Thus we do not expect the normalisation constant in the current two point function to be given by (\ref{C}) above. It should be given by a number
of the same order of magnitude, but the exact numerical factor is a strong coupling computation that could only be done if we knew the dual to QCD. 
We therefore leave the constant $C$ as a fitting parameter in what follows. The resulting value for $C$ is indeed of the correct order of magnitude, confirming our expectations.

\subsubsection{Polarization Basis\label{PB}}

To compute the cross section for  DVCS, and also for timelike deeply virtual Compton scattering  where an
outgoing timelike photon is emitted and then converted to a lepton or a $q\bar{q}$ pair, 
we need to contract the hadronic tensor with the photon polarization vectors.
Let $n_\lambda$ and $n'_{\lambda}$ be, respectively, the incoming and outgoing polarization vectors, normalized
such that $n^2=n'^2=1$ for transverse polarization, and $n^2=n'^2=\pm 1$ for longitudinal polarization
(time-like or space-like photon). 
We can use the gauge freedom to impose the conditions 
\begin{equation}
n \cdot k_1 =0\,,\ \ \ \ \ \ \ \ n' \cdot k_3 =0\,.
\end{equation}
Throughout the computation we work with spacelike off-shell photons and then take the on-shell limit, or analytically continue for the time-like case, if necessary.

In light-cone coordinates $(+,-,\perp)$, the polarization of the transverse incoming photon is 
\begin{equation}
n_\lambda = (0,0, \epsilon_\lambda)\,,\ \ \ \ \ \ \ \ \ \ \ \ \ \ \ (\lambda=1,2)
\end{equation}
where $\epsilon_\lambda$ is a orthogonal basis of unit vectors on $\bb{R}^2$. The incoming longitudinal photon has polarization
\begin{equation}
n_3 =   \frac{1}{Q}\left(\sqrt{s} , \frac{Q^2}{\sqrt{s}},0\right)\,,
\end{equation}
where we define $Q=\sqrt{Q^2}$. Note that  all the $n_i$ are orthogonal (and unit normalized).

For a transverse  outgoing photon (on-shell and off-shell) we have
\begin{equation}
n'_\lambda = \left( 0, 2\,\frac{\epsilon'_\lambda\cdot q_\perp}{\sqrt{s}} , \epsilon'_\lambda\right)\,,\ \ \ \ \ \ \ \ \ \ \ \ \ \ \ (\lambda=1,2)
\end{equation}
where  $\epsilon'_\lambda$ is a orthogonal  basis of unit vectors on $\bb{R}^2$.
For a longitudinal outgoing photon we have 
\begin{equation}
n'_3 =\frac{1}{Q'} \left(  \sqrt{s}, \frac{Q'^2+q_\perp^2}{\sqrt{s}} ,q_\perp \right)\,.
\end{equation}
Again all   $n'_i$ are orthogonal (and unit normalized).

\subsubsection{Deeply virtual Compton scattering}

We are finally in position to compute the DVCS cross section.
We need to square the amplitude, and then average over the incoming transverse 
photons and sum over final transverse on-shell photons.
For both incoming and outgoing transverse photons $(\lambda,\lambda'=1,2)$ a simple computation shows that
\begin{equation}
W_{\lambda \lambda'} = \left( n_\lambda\right)_a W^{ab}  \left( n'_{\lambda'} \right)^*_b =(\epsilon_\lambda\cdot \epsilon'^*_{\lambda'})  \, W_1=\delta_{\lambda\lambda'}  \, W_1\,,
\label{W_TT}
\end{equation}
where we define
\begin{equation}
W_n  = 2 i s QQ' \int dl_\perp\,
 e^{iq_\perp\cdot l_\perp} 
 \int \frac{dz}{z^3}\, \frac{d\bar{z}}{\bar{z}^3} \,
\Psi_n(z)\, \Phi (\bar{z})  \,
\left[ 1-e^{i\chi (S,L)}  \right]\,.\ \ \ \ \ \ \ \ (n=0,1)
\label{W_n}
\end{equation}
It is now clear why the limit $Q'\rightarrow 0$ of the amplitude $W_{\lambda \lambda'} $ is well defined and 
erases any IR dependence on the outgoing photon wave function, since when computing $W_1$ one takes
\begin{equation}
Q'   \left( K_1(Q' z) +\frac{K_0(Q'z_0)}{I_0(Q' z_0)}\,I_1(Q' z) \right) \rightarrow 1/z \,,
\end{equation}
where we are  accounting for the hard wall modification explained in (\ref{ProbeHardWall}). 
Finally we can square each amplitude, take the average over the incoming polarizations and sum over the final ones. 
We obtain 
\begin{align}
\frac{1}{2} \sum| W_{\lambda\lambda'} |^2 =  |W_1|^2 \,.
\label{W^2}
\end{align}

We remark that, in the longitudinal-transverse case we obtain after some algebra
that $W_{L \lambda'} =0$, so that helicity is conserved, as expected.

\subsubsection{Timelike deeply virtual Compton scattering}
\label{TDVCS}

Another interesting possibility is when the outgoing photon is timelike. It may then decay to a  lepton pair
($e\bar{e}$, $\mu\bar{\mu}$ or $\tau\bar{\tau}$), or to a $q\bar{q}$ pair, provided the photon momentum is
above the corresponding threshold. In particular, the process for lepton pair production is very clean and 
entirely within QED perturbation theory. Although we are not analysing in detail these processes in this paper, 
let us write for completeness the corresponding matrix elements involving longitudinal outgoing photons.  
In this case $Q'=\sqrt{Q'^2}$ is purely imaginary, so we need to be careful in analytically continuing the above expressions. 
Again helicity is conserved, so in addition to (\ref{W_TT}) we have the following non-vanishing
longitudinal-longitudinal matrix element
\begin{equation}
W_{LL} =  \left( n_3 \right)_a W^{ab}  \left( n'_{3} \right)^*_b = - W_0\,,
\end{equation}
where $W_0$ is also defined in (\ref {W_n}). It would be interesting to further explore this process using AdS/CFT and pomeron physics.

\section{$\gamma^*p \rightarrow \gamma p$ cross section }
\label{sigma}

Our goal is to predict the experimental data collected at HERA for the differential and total cross sections
in DVCS. For the differential cross section we have
\begin{equation}
\frac{d\sigma}{dt} (x,Q^2,t)= \frac{1}{16\pi s^2}\,\frac{1}{2} \sum| W_{\lambda\lambda'} |^2 = \frac{|W_1|^2}{16\pi s^2}   \,,
\label{differential}
\end{equation}
which depends on the three kinematical invariants $s$, $t$ and $Q^2$. For the available data the total cross section
\begin{equation}
\sigma (x,Q^2)=\frac{1}{16\pi s^2}  \int dt \, |W_1|^2  \,,
\label{total}
\end{equation}
is obtained from integrating the differential cross section from $0$ to  $-1\ {\rm GeV}^2$.

Let us then  further simplify the expression  for the function $W_1$ in (\ref{W_n}).
First we do the angular integral in the impact parameter space
\begin{equation}
W_n  \approx  4\pi i s\, QQ' \int_0^{z_0} \frac{dz}{z^3}\, \frac{d\bar{z}}{\bar{z}^3} \,
\Psi_n(z)\, \Phi (\bar{z})    \int_0^\infty dl \, l \, J_0(q l ) \left[ 1-e^{i\chi (S,L)}  \right] \,,
\end{equation}
where $q=|q_\perp| = \sqrt{-t}$ and $l=|l_\perp|$. 
Since we use a delta function for $\Phi(\bar{z})$ we can further simplify this expression to
\begin{equation}
W_n  \approx  4\pi i s\,QQ' \int_0^{z_0} \frac{dz}{z^3} \,
\Psi_n(z)  \int_0^\infty dl \, l \, J_0(q l ) \left[ 1-e^{i\chi (S,L)}  \right] \,,
\label{eq:W_n}
\end{equation}
where $L=L(z, z_*,l)$ defined in (\ref{S,L}) is now computed at $\bar{z} = z_*$.

We will now consider three models for the phase shift function $\chi (S,L)$: 
the AdS black disk, which is valid in a limited kinematical region of available data but very simple; the
conformal pomeron at strong coupling that neglects confinement corrections to the reggeon trajectory but is still rather simple and reproduces data in a very satisfactory way over a large kinematical range;  the hard wall pomeron that includes the effects of 
confinement in the previous model. 

\subsection{AdS Black disk}

We shall consider a  black disk model defined by a phase shift in the impact parameter representation of the hadronic tensor (\ref{HadronicTensor})
given by
$$
1- e^{i\chi(S,L)} = \Theta(L_s(S)-L)\,,
$$
where the saturation radius $L_s$ of the disk increases with energy  as
\begin{equation}
L_s(S) \approx \omega \log S\ .
\label{AdSdisk}
\end{equation}
This model reproduces quite successfully DIS data at low $x$, in the kinematical  region for which 
${\rm Im} \chi(S,L)\gtrsim 1$ and for a probe
above the QCD scale, i.e. for $Q \gg \Lambda_{QCD}$ \cite{Saturation,Cornalba:2010vk}. 
In this simple model the phase shift is characterized by the single parameter $\omega>0$.
As long as the imaginary part of the phase shift grows as $L$ decreases, one expects this model to be a good approximation
provided the external kinematics imply that the dominant region of integration in the holographic variables $z$ and $\bar{z}$ 
is such that $L<L_s$. This model can be motivated using the BFKL pomeron at weak coupling, 
the gravi-reggeon and the phenomena of geometric scaling observed in DIS at low $x$, as explained in \cite{Saturation,Cornalba:2010vk}. 

\begin{figure}[tbp]
\begin{center}
\includegraphics[scale=0.4]{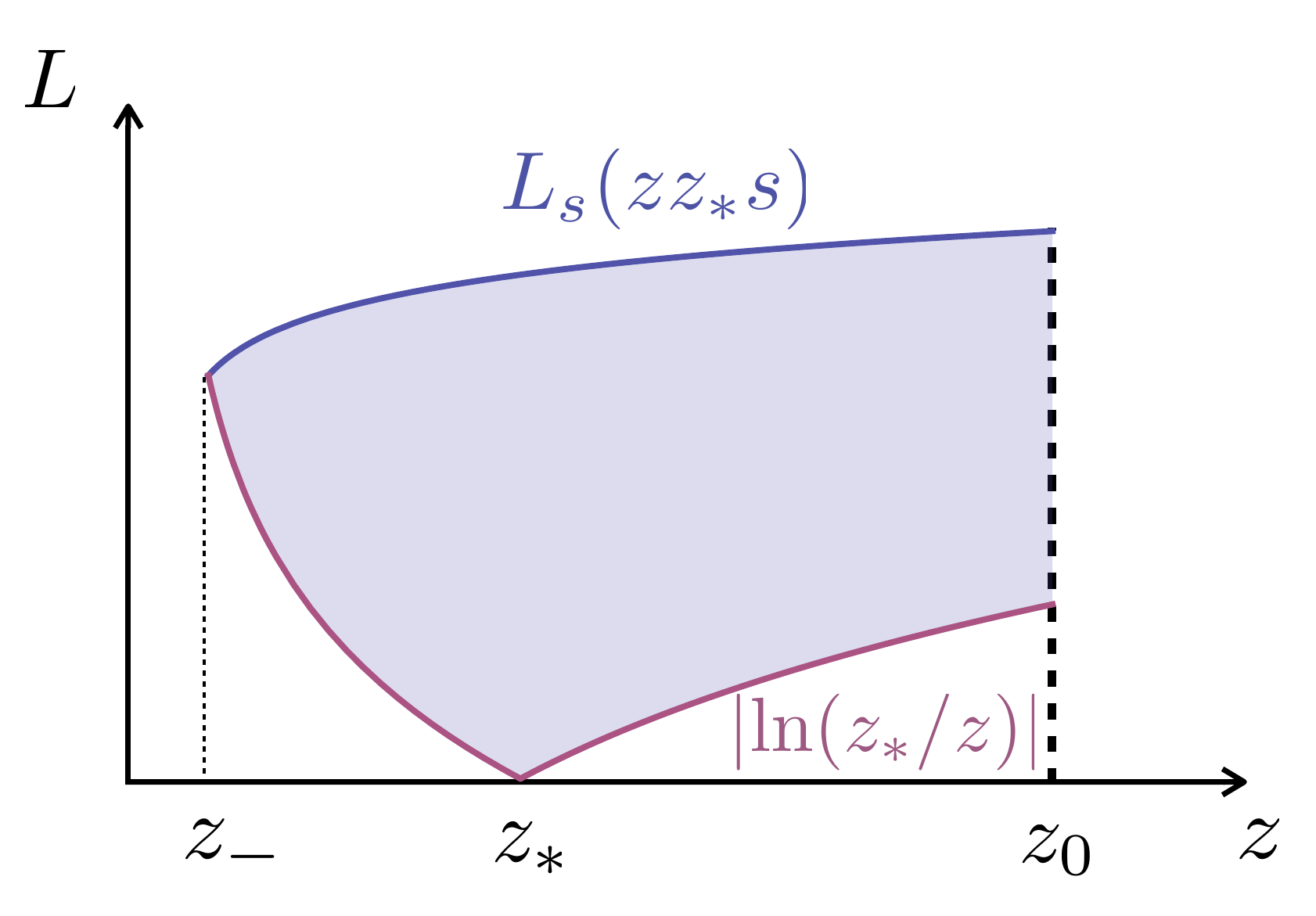}
\caption{{\small The upper and lower bounds of integration in the $H_3$ impact parameter $L$. Since the probe wave function is localized near
 the boundary, the dominant region of integration is within the interval $z_-<z<z_*$.}}
\label{Lintegration}
\end{center}
\end{figure}

For the black disk model, the integral over the impact parameter can be better done by moving to the AdS impact parameter $L$ where
the phase shift is trivial. However one needs to be careful with the limits of integration. For fixed $z$,
the upper limit of integration in $L$ is determined by the size of the black disk $L=L_s(zz_*s)$. 
The lower limit is determined by setting $l_\perp=0$, i.e. for $L=\left|  \ln (z_*/z) \right|$.
This region of integration, as a function of $z$, is shown in figure  \ref{Lintegration}. 
However when changing integration variables $(z,l)$ to $(z,L)$ the region of integration in $z$ is also bounded by the condition
$L_s > \left|  \ln (z_*/z) \right|$, which introduces a  lower bound of integration in $z$. 
The lower bound $z_-$ is simple to understand from the AdS perspective. For a fixed target at $z_*$, as the probe gets close to the 
boundary, the AdS impact parameter grows until it reaches the boundary of the disk for $L(z,z_*,0)=L_s(zz_*s)$. Thus the condition defining $z_-$ is given by
\begin{equation}
  \frac{z_-}{z_*} + \frac{z_*}{z_-} = (z_- z_* s)^{\omega} + (z_- z_* s)^{-\omega} \,.
\end{equation}
There are two solution of this equation, obtained one from another by $\omega  \rightarrow -\omega$. We are interested in the solution\footnote{The other solution
gives an upper bound for the $z$ integration, corresponding to the probe  placed deep in the infrared with respect to the target. We do not need
 to take this into account because the cut-off  $z_0$ will be close enough to $z_*$. Also, since the probe wave function will be localized near the boundary in the UV,
the result will in practice be independent of $z_0$.}
\begin{equation}
 (z_- z_* s)^{\omega}  = \frac{z_*}{z_-}  \ \ \ \Rightarrow  \ \ \   z_-^{1+\omega} = z_* (z_* s)^{-\omega}\,.
\end{equation}
This lower bound is also shown in  figure  \ref{Lintegration}. 
We conclude that
\begin{equation}
W_n  \approx  4\pi i s QQ'  z_*  \int_{z_-}^{z_0} \frac{dz}{z^2}\,
\Psi_n(z)\,   \int_{\left|  \ln (z_*/z) \right|}^{L_s(zz_*s)} dL \sinh L  \, J_0(q l) \,,
\end{equation}
where the function  
\begin{equation}
l=l(L,z,z_*) =\sqrt{2zz_*\cosh L - z^2 -z_*^2}
\end{equation}
is trivially obtained from (\ref{S,L}). 
The $L$ integral can be done by writing $u= (q l)^2$, with the result
\begin{equation}
W_n  \approx  4\pi i \,\frac{s}{q} \,QQ' \int_{z_-}^{z_0} \frac{dz}{z^3}\,
\Psi_n(z)\,l_s\, J_1(ql_s)\,,
\end{equation}
where $l_s = l(L_s,z,z_*)$. Taking the $Q'\rightarrow 0$ limit, and replacing for the expression of $\Psi_1$
we finally obtain
\begin{equation}
W_1 \approx-  C\, \frac{    2\pi^{3} }{3  } \,i\,\frac{s}{q} \,Q\int^{z_0}_{z_-}\, dz\, K_1(Q z)\, l_s\, J_1(ql_s)\,.
\label{BDfinal}
\end{equation}
This expression depends on the three parameters  $C$,  $\omega$ and $z_*$. In general it also depends on $z_0$, but for a probe localized in the 
UV region, this dependence drops out. In the next section we will use this expression in the
differential cross section  (\ref{differential}) and in the total cross section  (\ref{total})
to compare with HERA data.

\subsection{Conformal pomeron}

The next model we will consider for $\chi(S,L)$ is a model for pomeron exchange in a conformal theory. 
In the case of an amplitude dominated by single pomeron exchange, the phase shift $ \chi= \chi(S,L)$  is small and is related to the 
conformal amplitude introduced in section \ref{CFTRegge}  by ${\cal B} =  1- e^{i \chi} \approx - i \chi$. 
Explicitly our single conformal pomeron exchange model  is obtained by using expression (\ref{W_n}) 
for $W_1$ with the conformal amplitude computed in (\ref{AmplitudeFinal}).

Our conformal pomeron exchange model depends on three parameters. One of them is the position of the target $z_*$, 
and is the same as in the previously discussed black disk model. The next parameter is $\rho$, related to the pomeron intercept via $\rho=2-j_0$.
In terms of the 't Hooft coupling, defined by the ratio  $R^4/\alpha'^2$,
it is given by $\rho\, = \, 2/\hksqrt{\lambda}$.
Finally, the amplitude will also depend on the overall  constant  $C g_{0}^{2}$, where $C$ is given by the normalization of the current two point function and $g_0^2$ by the 
impact factors of the external states. It is clear that for the single pomeron exchange we can not determine both constants independently.

We expect the conformal pomeron model to give good results in the limit of $Q^2$ large, with $x$ small, but not too small. We need $x$ to be small in order to be in the kinematical regime where the pomeron exchange dominates. However, taking the limit of $x\rightarrow 0$ of equation (\ref{AmplitudeFinal}), we can see that its asymptotic behavior for the cross section
is
\begin{equation}
\sigma\sim (1/x)^{2(j_0-1)} \gg \Lambda_{QCD}^{-2} \ln^2(1/x) \,,
\end{equation}
which violates the Froissart bound. This limit would also make ${\rm Im}\, \chi \gg 1$, which means that at very small $x$ (\ref{AmplitudeFinal}) is
no longer valid, and we expect multi-pomeron exchange effects to play an important part.  If we perform the eikonal resummation of
pomeron exchanges our amplitude then becomes the black disk model of the previous section, as shown by figure \ref{BlackDisk} of section \ref{BD}. 
However, the black disk amplitude (\ref{BDfinal}) also violates the Froissart bound. This is not surprising, and it is not a contradiction,
because both models are constructed from a conformal amplitude $B(S,L) $, without a scale to define the Froissart bound in the first place.

To better understand the amplitude  as we go to very  small $x$ at fixed $Q^2$, we need to realise that confinement  effects also start to 
play a crucial role, as shown in \cite{Brower:2010wf}, and further discussed in section \ref{sec:HWP} below. Moreover, when $Q^2$ or the 
momentum exchange $t$ approach the QCD scale, we also expect confinement  effects
to become important.  This brings us to the hard wall pomeron model that we now discuss.

\subsection{Hard wall pomeron}\label{sec:HWP}

The  hard-wall model is obtained by placing a sharp cut-off on the radial AdS coordinate at $z=z_0$. This changes the boundary conditions of the differential equations, and hence leads to novel results. The advantage of this model is that it is  relatively straightforward 
to implement in AdS/CFT, leading to equations that can often still be solved analytically. The cut-off also sets a scale in the theory, and therefore breaks conformal invariance and leads to a theory of confinement. Far away from the cut-off, its effects should be only weakly felt though, and we would recover once again approximate conformality, as in QCD. To see intuitively why the hard-wall model is confining, we can start by thinking about how the extra coordinate arises in the AdS/CFT correspondence in the first place. We can think of the graviton on the AdS side as being a product of two gluons from the CFT side. The radial coordinate in such a picture corresponds to the separation between the two gluons \cite{Polchinski:2010hw}. Therefore putting a maximum distance on this coordinate gives us a maximum separation between the gluons, and hence confinement.

Let us commence by looking more closely at the relationship between the pomeron in the conformal and the hard-wall models. As mentioned before, the pomeron propagator in both these models satisfies the same differential equation, but with different boundary conditions \cite{Brower1}. First notice that at $t = 0$, equation  (\ref{AmplitudeFinal}) can be integrated 
in impact parameter $l_\perp$, with the result
\begin{align}
\chi(\tau,t=0,z,\bar{z}) &= 2\pi z\bar{z} \int_{\left| \ln(\bar{z}/z)\right|}^\infty dL\,\sinh L\, \chi(\tau,L)\nonumber
\\
 & =   i \pi \,g_0^2  \left(  \cot\Big(\frac{\pi \rho}{2}\Big) + i \right)  (z\bar{z})
\, e^{(1-\rho) \tau}\,  \frac{e^{-\frac{(\ln(\bar{z}/z))^2}{\rho\tau}}}{  (\rho \tau)^{1/2}}\,,
\end{align}
where we defined the rapidity $\tau = \ln(\alpha' S)$.
Similarly, the $t = 0$ result for the hard-wall model can also be written explicitly
\begin{equation}
  \label{eq:chihw_t0}
  \chi_{hw}(\tau,t=0,z,\bar{z})=\chi(\tau,0,z,\bar{z})+{\cal F}(\tau,z,\bar{z})\, \chi(\tau,0,z,z_0^2/\bar{z})\, .
\end{equation}
We see that at $t = 0$ the hard-wall result can be represented as a sum of two conformal amplitudes, 
with the second one representing a conformal kernel on the other side of the hard-wall. The function 
\begin{equation}
  \label{eq:F}
  \mathcal{F}(\tau,z,\bar{z}) = 1-4\hksqrt{\pi\tau}\,e^{\eta^2}\erfc(\eta)\,, \ \ \ \ \ \ \  \eta = \frac{-\log(z\bar{z}/z_0^2)+4\tau}{\hksqrt{4\tau}}\,
\end{equation}
is set by the boundary conditions at the wall and represents the relative importance of the two terms and therefore confinement. This function varies between $-1$ and $1$, approaching $-1$ at either large $z$, which roughly corresponds to small $Q^2$, or at large $\tau$ corresponding to small $x$. It is therefore in these regions that confinement is important. This relationship has been explored before in \cite{Brower:2010wf}. Since we are working at small $x$, the size of ${\cal F}$ will roughly vary between $-0.1$ and  $-0.4$
for the data here analysed. 

To use the hard-wall propagator in equation (\ref{eq:W_n}) we need to have the propagator at finite impact parameter $l$. Work is in progress in evaluating it numerically. In this paper, we used the same approximation as in \cite{Brower:2010wf}. It can be shown that at large $l$ the eikonal for the hard-wall model has a cut-off
\begin{equation}
  \label{eq:hw_eik}
   \chi_{hw}(\tau ,l,z,\bar{z})\sim \exp[-  m_1 l  - (m_0-  m_1)^2\;  l^2 / 4 \rho \tau ]\,,
\end{equation}
where $m_1$ and $m_0$ are  solutions of \footnote{ $m_1 \approx  1.6/z_0$ and  $m_0 \approx 3.8/z_0$ are the masses of the lightest spin 2 glueballs on the lowest two Regge trajectories. For more on calculating glueball masses see \cite{Brower:2008cy}.}
\begin{equation}
 \left.  \partial_z(z^2J_0(m_1z))\right|_{z=z_0} =0\,,\ \ \ \ \ \ \ \ \ 
\left.  \partial_z(z^2J_2(m_0z))\right|_{z=z_0} =0\,.
\end{equation}
At small $l$ we assume $\chi$ still has the $t=0$ form of a sum of two conformal kernels, but now with $l$ dependence as well
\begin{equation}
  \chi^{(0)}_{hw}(\tau,l,z,\bar{z})\sim  \chi_{c}(\tau ,l,z,\bar{z}) + {\cal F}(\tau ,z,\bar{z}) \,\chi_{c}(\tau,l,z,z_0^2/\bar{z}).
\end{equation}
Finally we will introduce a normalization function, $C(\tau,z,z')$, independent of $l$, which ensures that the result (\ref{eq:chihw_t0}) is reproduced after integrating over $l$. 
Putting it together, the approximation we use for $\chi$ in the hard-wall model is
\begin{equation}
  \label{eq:chi_hw}
  \chi_{hw}(\tau,l,z,\bar{z}) = C(\tau,z,\bar{z})\,D(\tau ,l)\, \chi^{(0)}_{hw}(\tau,l,z,\bar{z}),
\end{equation}
where   
\begin{equation}
  \label{eq:D}
  D(\tau,l) = \min\left(1,\frac{ \exp[-  m_1 l  -  (m_0-m_1)^2 l^2 / 4 \rho \tau ]}{\exp[-  m_1 z_0  - ( m_0-m_1)^2  z_0^2 / 4 \rho \tau ]}    \right)
\end{equation}
is the exponential cutoff at large $l$. We expect this to be a good approximation in the range of $(x,t,Q^2)$ where experimental data is currently available. This model has one more parameter, $z_0$, than the conformal pomeron model.\footnote{Note that even though $z_0$ appears in the equation for the conformal model as well, it is not really a parameter there since, due to the shape of the probe wave function, as long as $z_0 \gtrsim 5\  {\rm GeV}^{-1}$ the result will be independent of its value.}

\subsection{Eikonalization}\label{sub_sec:gamma:eikonal}

As mentioned in the previous two sections, at very small $x$ single pomeron exchange violates the Froissart bound, which is common to both the conformal and the hard-wall models. One way to avoid this difficulty is to fit directly equation (\ref{eq:W_n}) to data, without expanding the exponent in $e^{i\chi}$, where $\chi$ is taken as the amplitude for 
single pomeron exchange. This would introduce an extra parameter in the fit, the overall normalization factor. Staying at single pomeron exchange level, only the product of this parameter with $g_0^2$ appears, and so it cannot be fit. However, in the kinematical region where data is available, the size of $\chi$ is small, and eikonalization does not change the value of our parameters, nor improve the quality of the fit. This is consistent with the conclusions of \cite{Brower:2010wf}. As data at smaller values of $x$, with $Q^2$ fixed, becomes available, we expect eikonalization to play a more important part, and to enter the black disk regime.

That being said, we will briefly revisit eikonalization in section \ref{sub_sec:data:eikonal}. Unlike for single pomeron exchange, $C$ and $g_0^2$ can be fit separately in the eikonal model, since $g_0^2$ is exponentiated and $C$ is not. This will therefore  allow us to estimate the size of $C$ and to compare it with that obtained in the black disk and with the weak coupling result
(\ref{C}).

It is also worth noting that, in contrast with the conformal models,  the hard-wall $\chi$ given in equation (\ref{eq:chi_hw}) will satisfy the Froissart bound, due to the exponential cut-off, (\ref{eq:D}).

\section{Data analysis} \label{Data}

\begin{figure}[htbp]
\begin{center}
\includegraphics[scale=0.44]{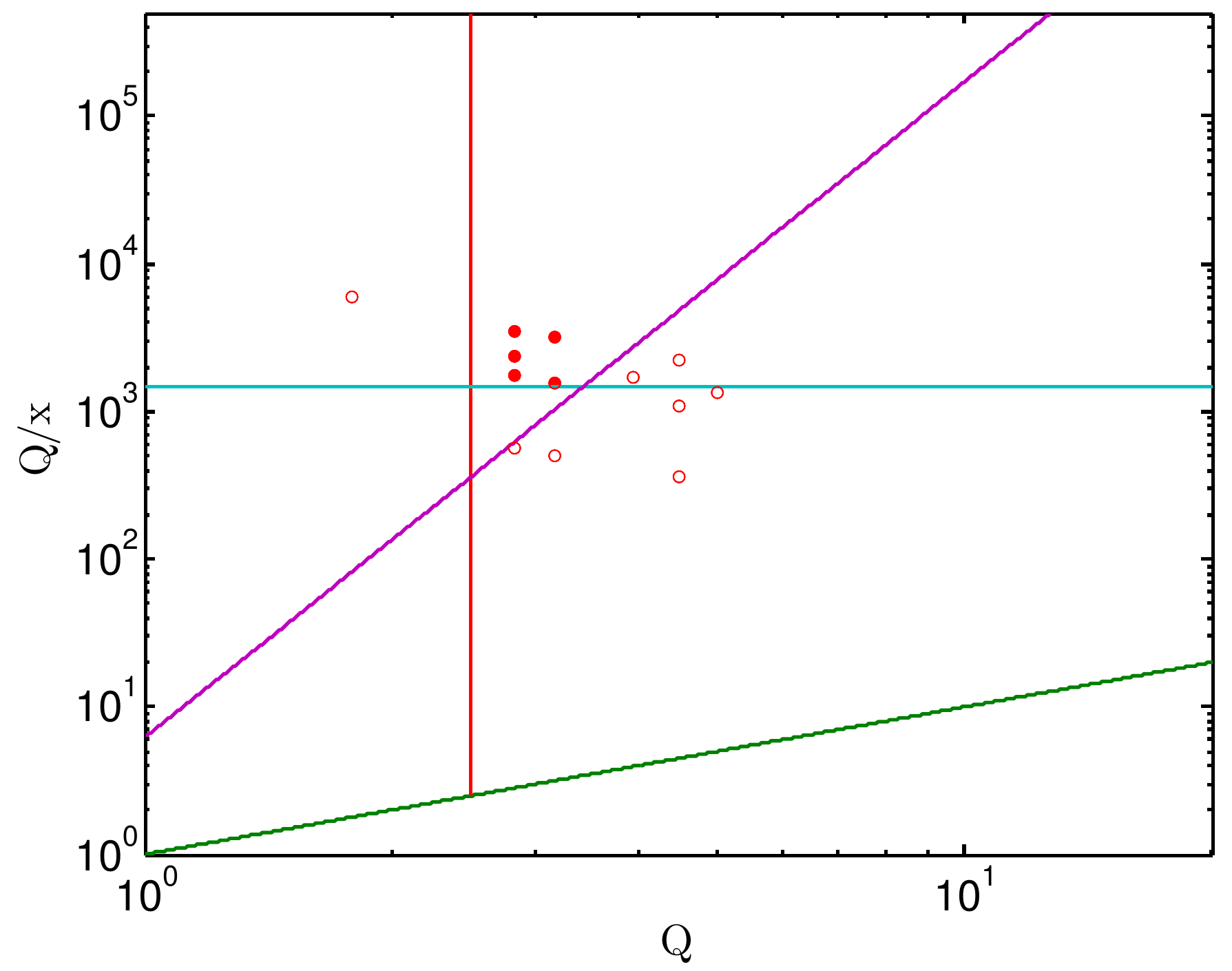}
\includegraphics[scale=0.44]{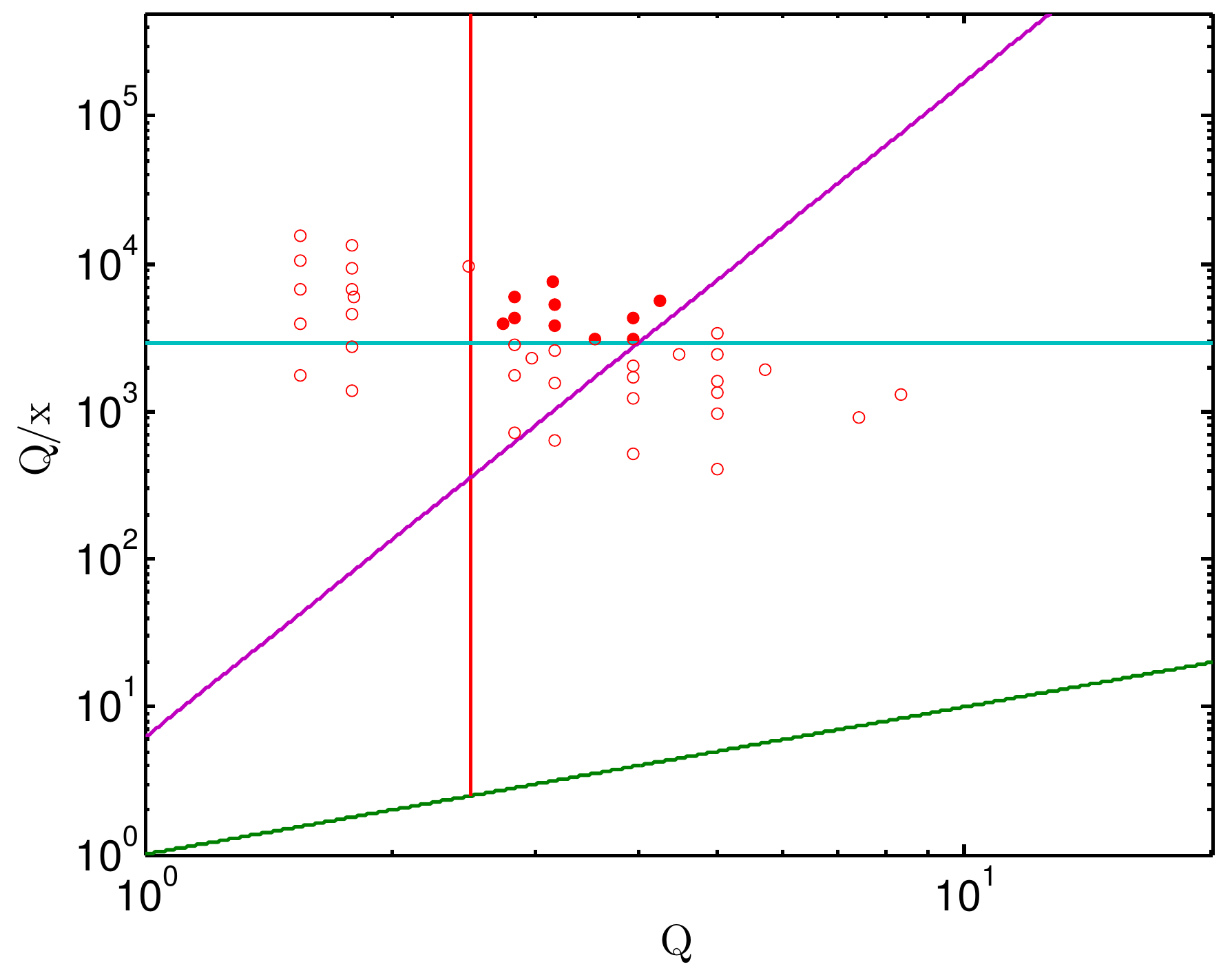}
\caption{{\small (Left) Data points to fit  the differential cross section. Note that for each point we have 4 different values of $-t$, so we have a total of 52  points. In the case of the black disk model
we restrict the analysis to the filled dots (20  points), as explained in the text. (Right) Data points to fit  the total cross section, giving a total of 44 points. 
The filled dots represent the points we select to fit the cross section using the black disk model (10  points). In both figures $Q$ is in ${\rm GeV}$.}}
\label{fig:ranges}
\end{center}
\end{figure}

We will now compare our results to data collected at HERA. We will use the data from both ZEUS and H1 experiments, taken from their latest publications on DVCS \cite{Chekanov:2008vy,:2009vda}. The data sets have results for both the differential cross section (\ref{differential}), and the cross section (\ref{total}). 
Figure \ref{fig:ranges} shows all the data points analysed in this paper, as a function of $S\sim Q/x$ and $Q$.
Compared to DIS, the data sets are smaller, with $52$ and $44$ points respectively. In a subsequent publication, we plan to extend our study to vector meson production as well, which will give us more data. 

%
%

In turn, we will look at each of our models. We will see that the fits give us precisely what we expect from the theory. In the case of the differential cross section, the best fit is obtained using the hard-wall model. The conformal pomeron model begins to show signs of breaking down at small values of $t$, which we expect to be the case due to the effects of confinement, which is precisely what the hard-wall model incorporates. The black disk model gives a very good agreement with experiment in a limited regime where we a priori expect it to work - when the size of $\chi$ starts increasing, but before we reach the region where confinement is important. 

For the cross section, the values of $x$ and $Q^2$ are in the region where we expect the conformal model to work, and before confinement starts becoming important. This is borne out in the fits, with both the conformal and hard wall models giving very good fits. The conformal fit is slightly better, but the difference is not large. Once again, the black disk model also gives a very good fit in the  region where we expect it to work. 

Note that the normalization of the photon wavefunction $C$ is a parameter in all models, but for single pomeron exchange it only appears in the combination $C g_0^2$ and hence these two parameters cannot be fit separately. The results we present below in sections \ref{sub_sec:data:conf_pomeron} and \ref{HWP} were obtained by fixing $C$ to the 
reference value (\ref{C}), given by the weak coupling computation. 
Therefore bear in mind that even though we quote a value for $g_0^2$ it should be understood that we only know the value of the product $C g_0^2$.

\subsection{Black disk}

The black disk model is expected to be valid in a kinematical window where QCD is approximately conformal and where the 
amplitude for single pomeron exchange becomes large. This reduces the number of available data points to a small subset. 
The criteria to select data points was discussed in detail in the original work \cite{Saturation}, here we only summarise it. 
There are three conditions we need
to impose: $(i)$ $Q>Q_{min} \sim 2\  {\rm GeV}$, so that our probe does not feel confinement effects; $(ii)$ To be inside the saturation
region, defined by the condition ${\rm Im}\, \chi \sim 1$, we need to impose   $\omega \ln(z_* Q/x) > \ln (z_* Q)$, which can be 
derived from the expression of the conformal amplitude for the exchange of a reggeon; $(iii)$ We need to be in the Regge limit of large
$s$ which, given the value of the other kinematic invariants,  gives the condition $z_* Q/x > 10^3$. The straight lines
plotted in figure \ref{fig:ranges} correspond to these three conditions, so that we are left with 20 points for the differential cross section 
and 10 points for the cross section.

The parameters we get from the differential cross section data are\footnote{Throughout the paper we fit using a weighted non-linear least-squares method, and the parameter errors represent the 95 percent confidence interval for the parameter estimates.}
\begin{equation}
\label{eq:dsdt_bd_pars}
  \omega = 0.243 \pm 0.045\,,\ \ \ \ z_* = 3.50 \pm 0.89\ {\rm GeV}^{-1}\,,\ \ \ \ C = 0.0013 \pm 0.0010\, ,
\end{equation}
which gives us a $\chi^2$ of
\begin{equation}
  \chi^2_{d.o.f.} = 0.97\, ,
\end{equation}
showing that the black disk fits in the range we expect. Analogously, the parameters from the total cross section data are
\begin{equation}
  \omega = 0.275 \pm 1.581\,,\ \ \ \ z_* = 1.39 \pm 124.80\ {\rm GeV}^{-1}\,,\ \ \ \ C = 0.00094 \pm 0.00530 \,,
\end{equation}
corresponding to a $\chi^2$ of
\begin{equation}
  \chi^2_{d.o.f.} = 1.22\,.
\end{equation}
  
\begin{figure}[htbp]
\begin{center}
\includegraphics[scale=0.435]{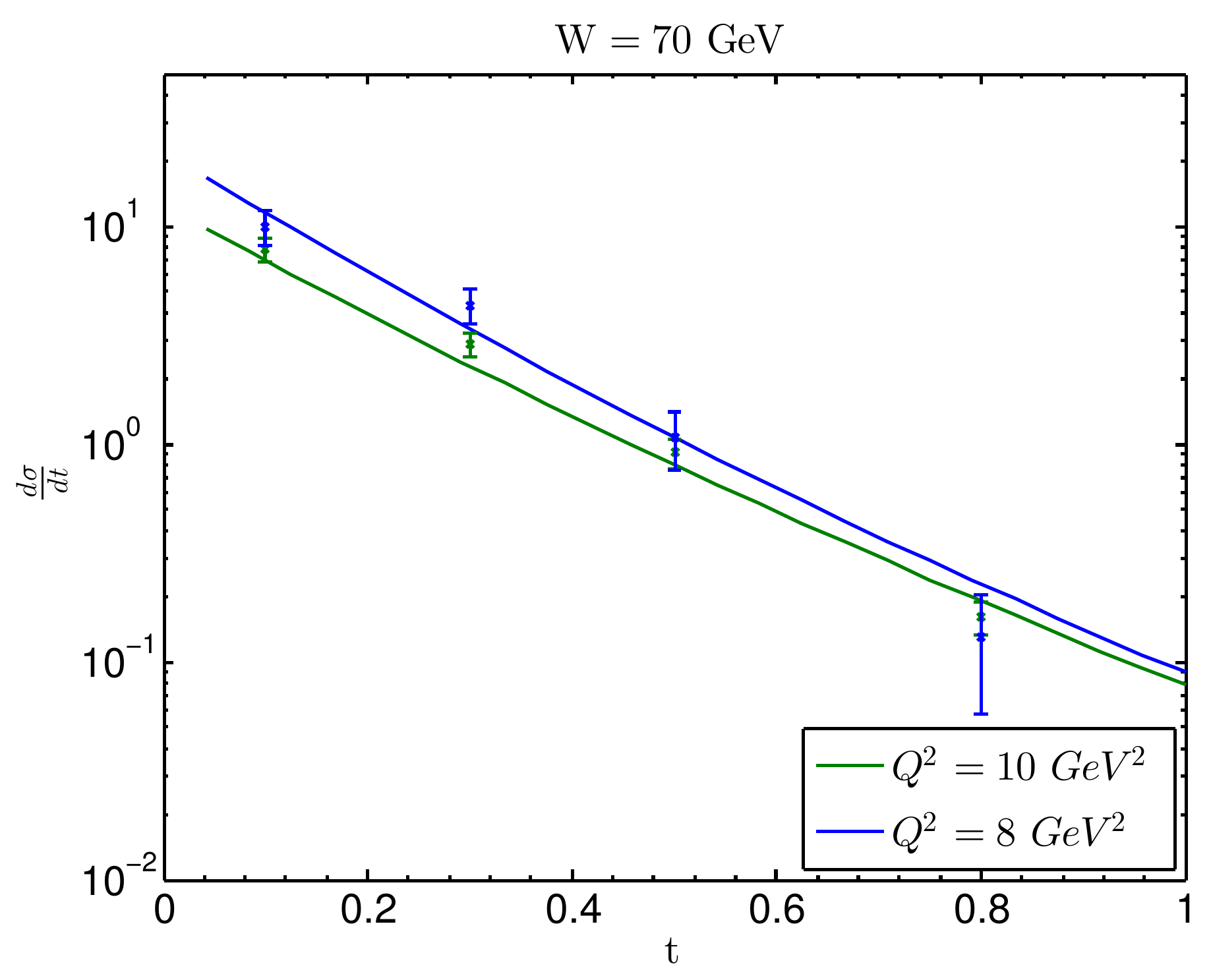}
\includegraphics[scale=0.435]{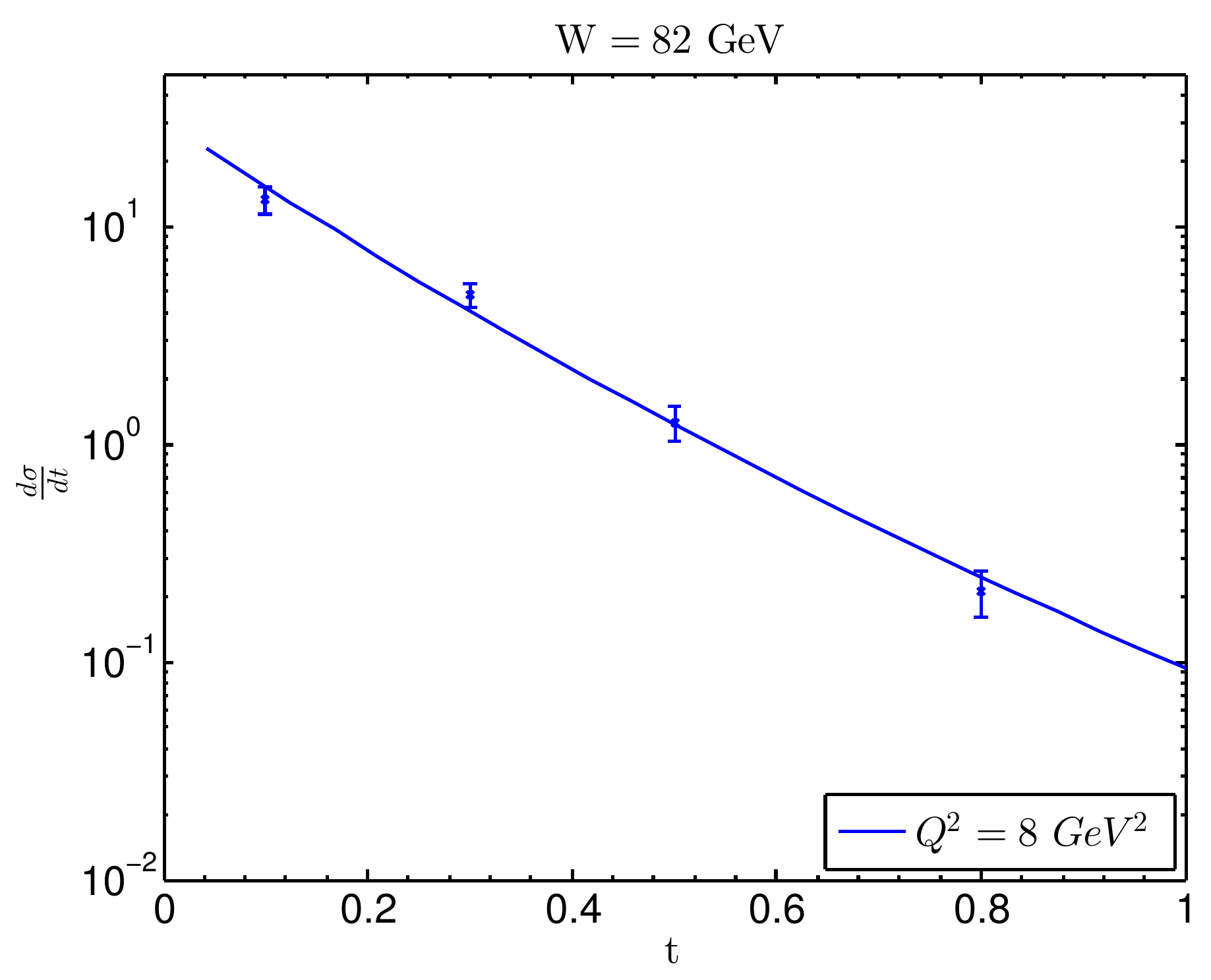}\\
\includegraphics[scale=0.435]{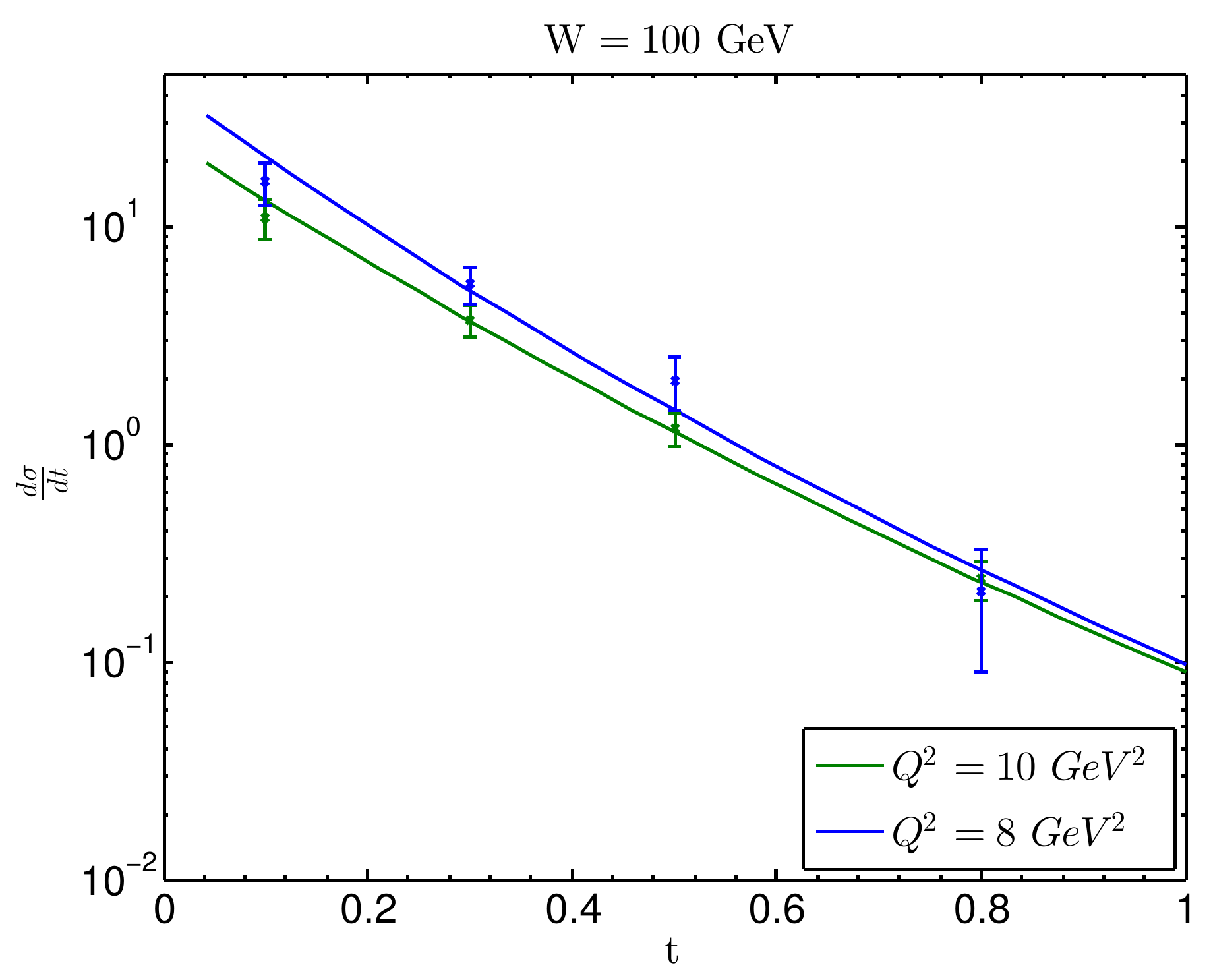}
\includegraphics[scale=0.435]{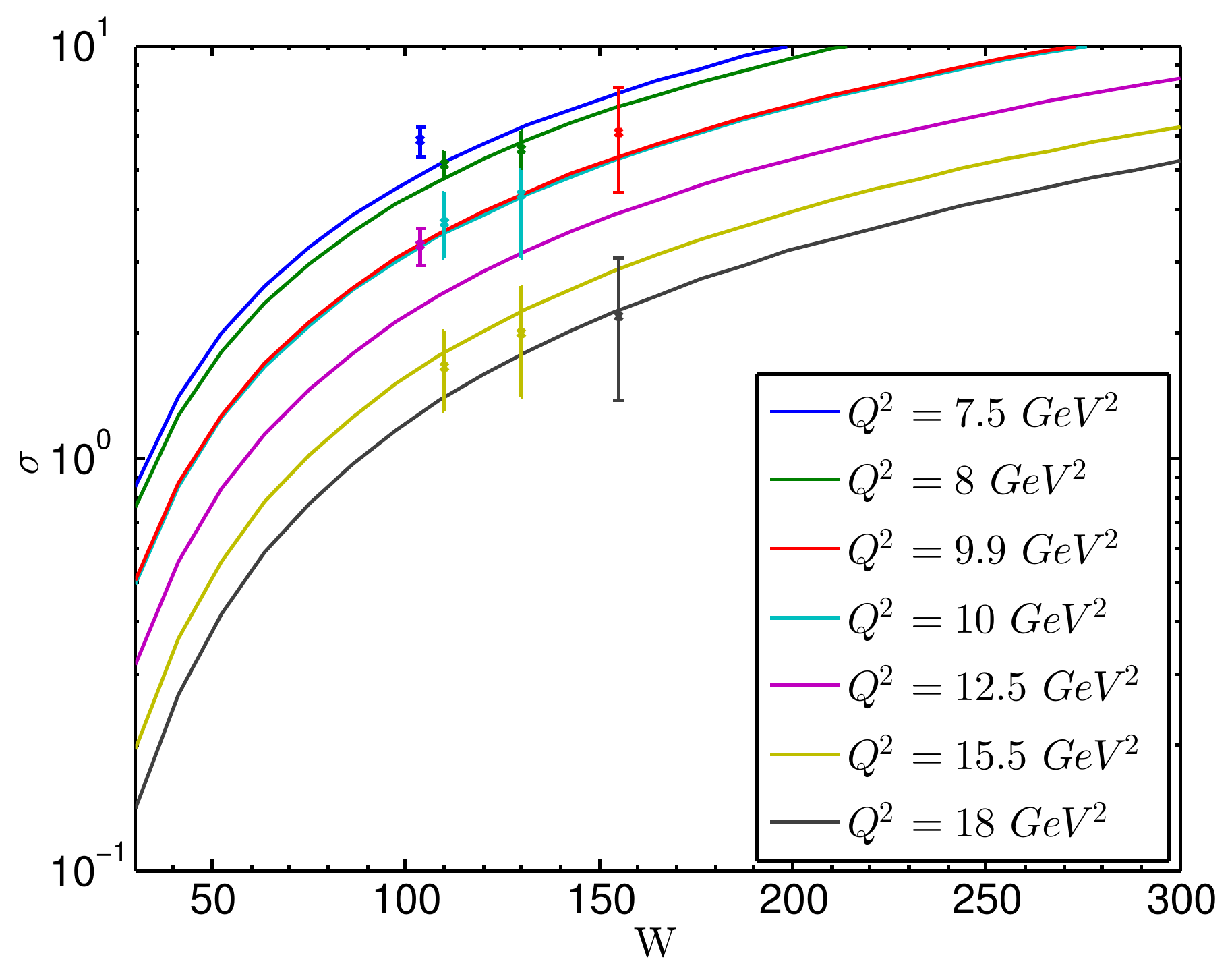}
\caption{{\small The fit of the black disk model to the data. The first three figure are for the differential cross section and the last one for the cross section.
In these figures $W$ is the center of mass energy, $W=\sqrt{s}$, and we use units of ${\rm GeV}$.}}
\label{fig:bd_all}
\end{center}
\end{figure}

In figure \ref{fig:bd_all} we see the plots for these fits. Considering that we are fitting a very small number of data points, the results are very good.
Both values for the exponent $\omega$ are very close, and correspond to an intercept $j_0=1+\omega$ between $1.2$ and $1.3$, as expected. 
Also, for the overall constant $C$ and  the proton scale $z_*$ we obtain values with the expected order of magnitude, as given by the weak coupling result 
(\ref{C}) and the proton mass, respectively. Although results are making sense, with very good values for $\chi^2$, 
a word of caution is required for the very large errors obtained in the estimated parameters. This happens because we have a very small number of data points
in a very limited region where the black disk model only starts to describe the physics. With data for smaller values of $x$, inside the region defined by the black disk criteria
plotted in figure \ref{fig:ranges}, we would expect to be able to considerably reduce these errors.

%
\subsection{Conformal pomeron}\label{sub_sec:data:conf_pomeron}

\begin{figure}[htbp]
\begin{center}
\includegraphics[scale=0.43]{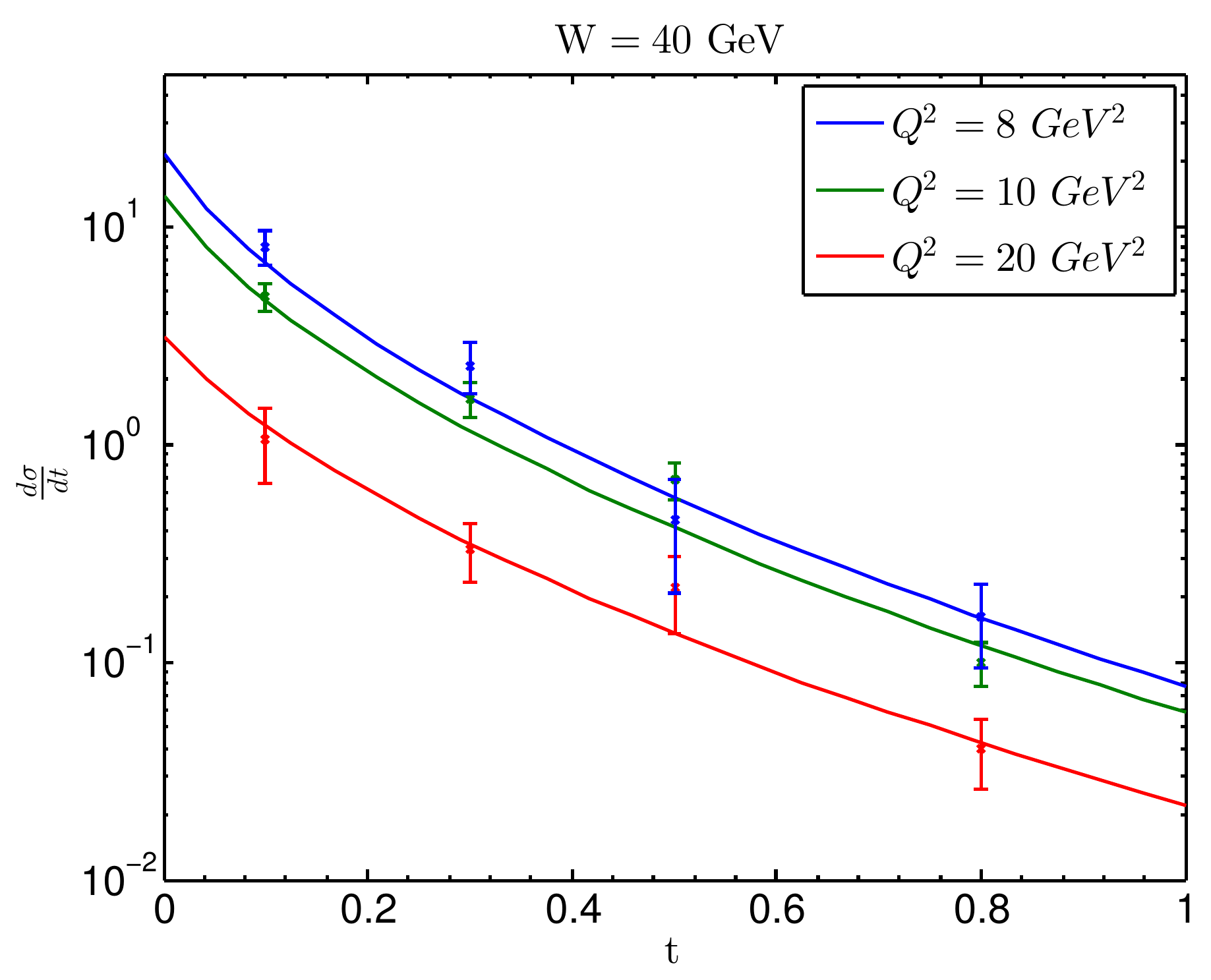}
\includegraphics[scale=0.43]{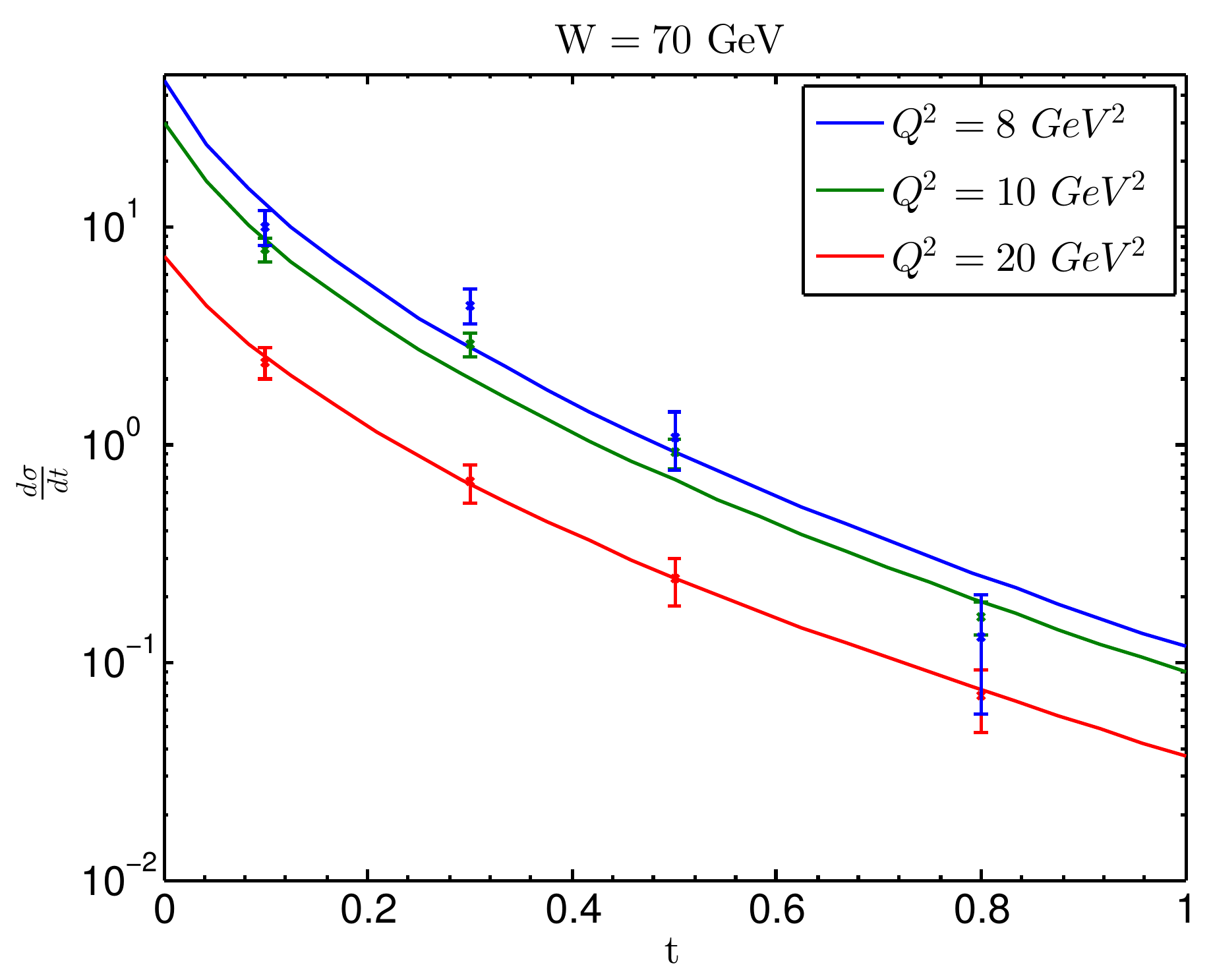}\\
\includegraphics[scale=0.43]{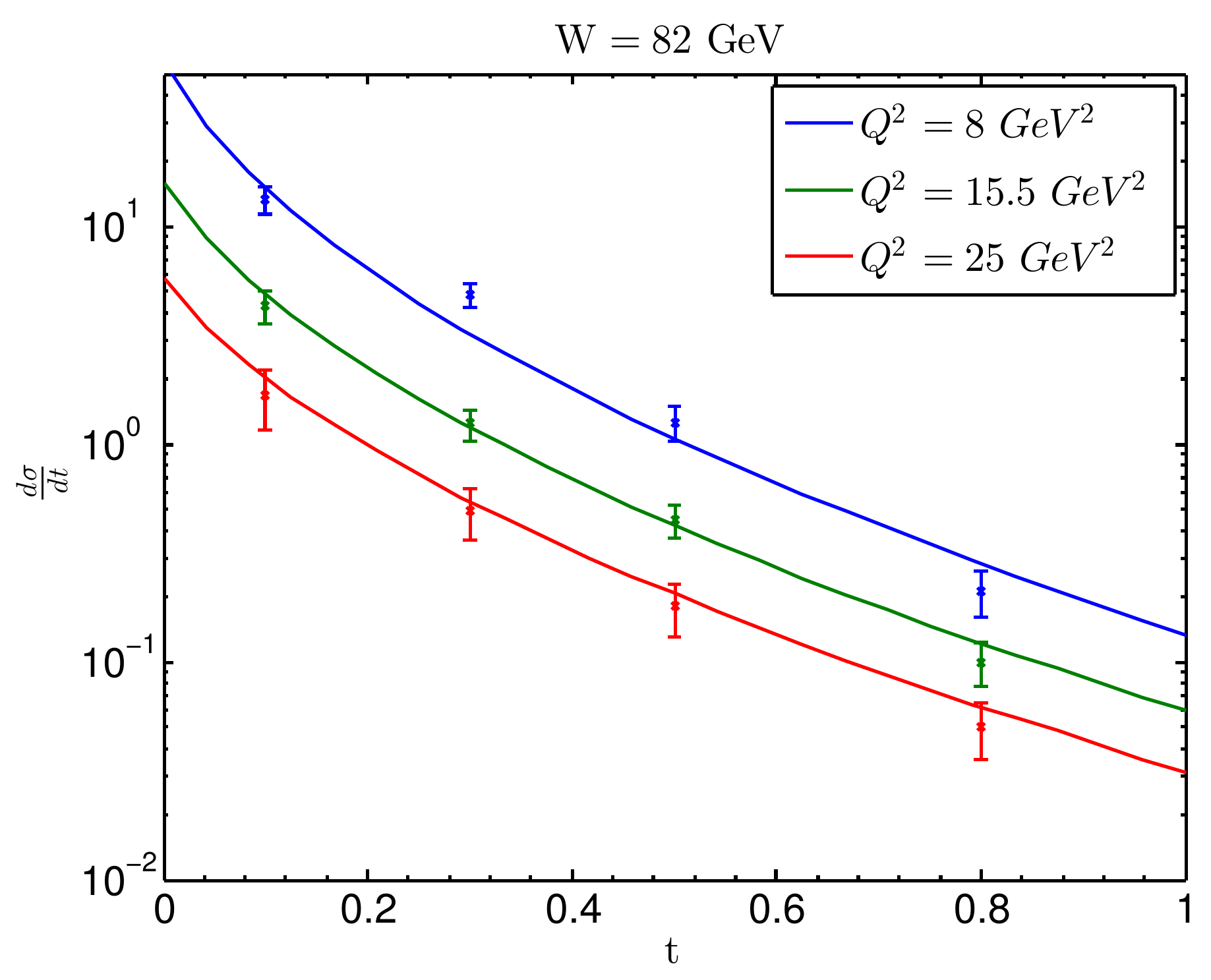}
\includegraphics[scale=0.43]{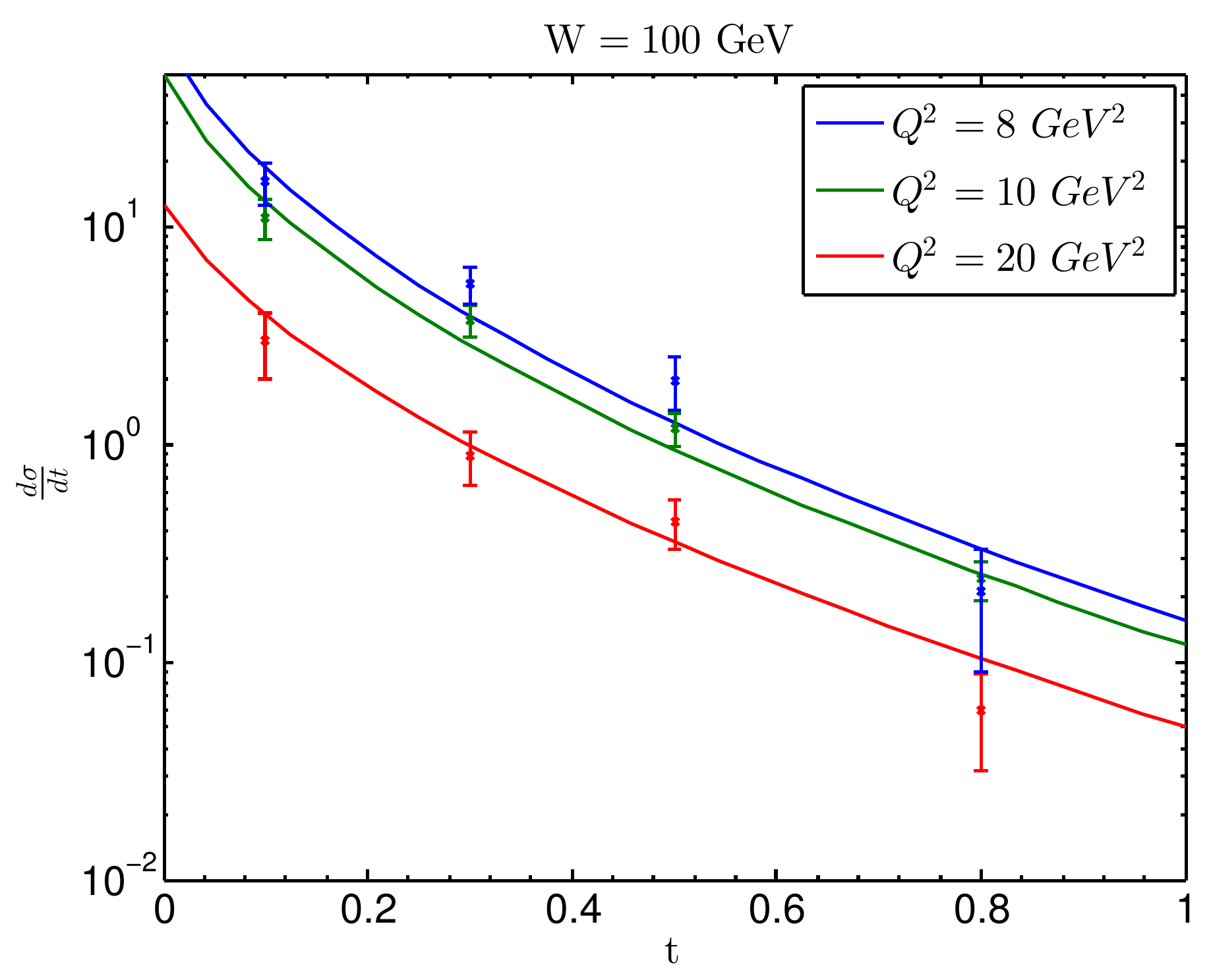}\\
\includegraphics[scale=0.43]{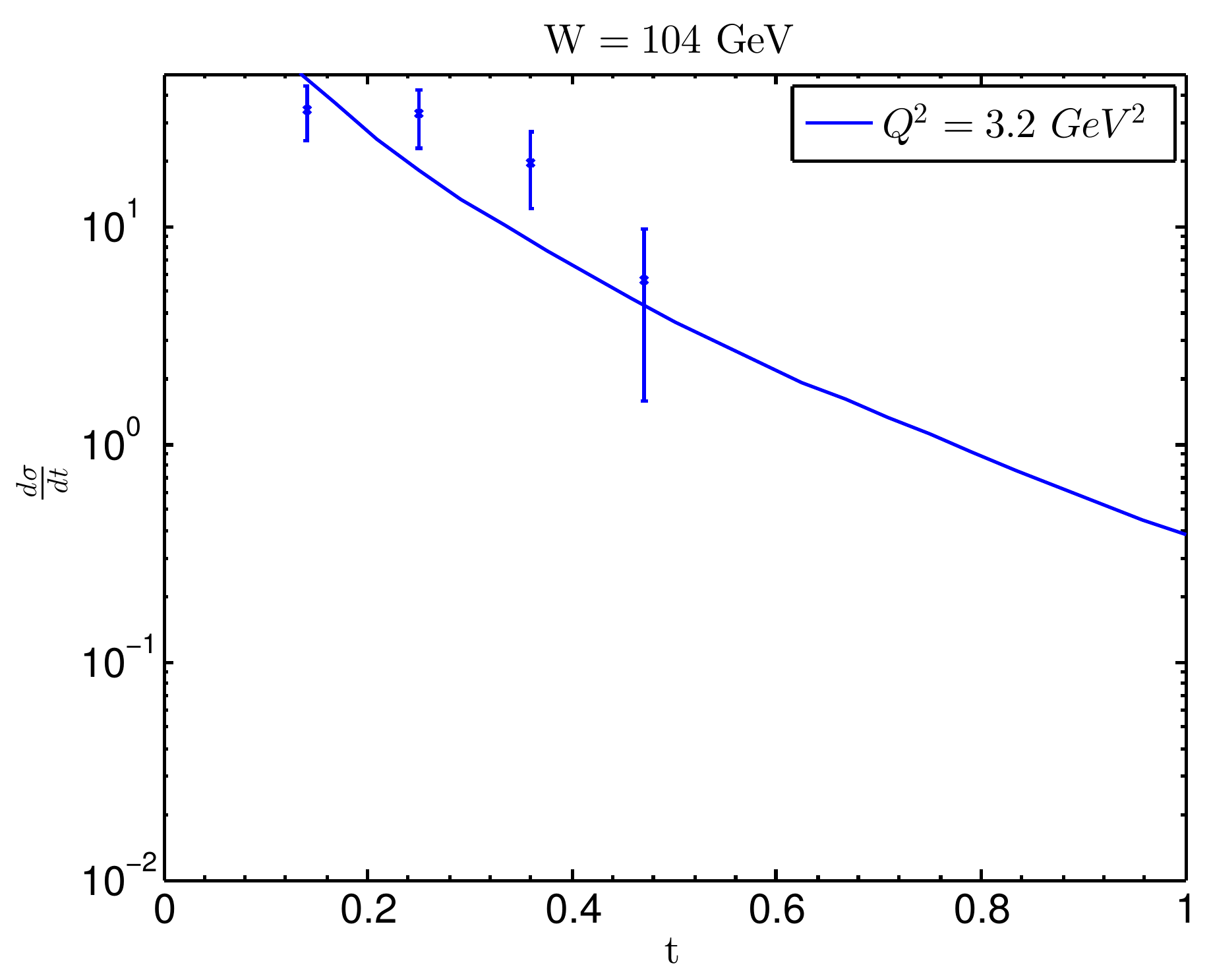}
\includegraphics[scale=0.43]{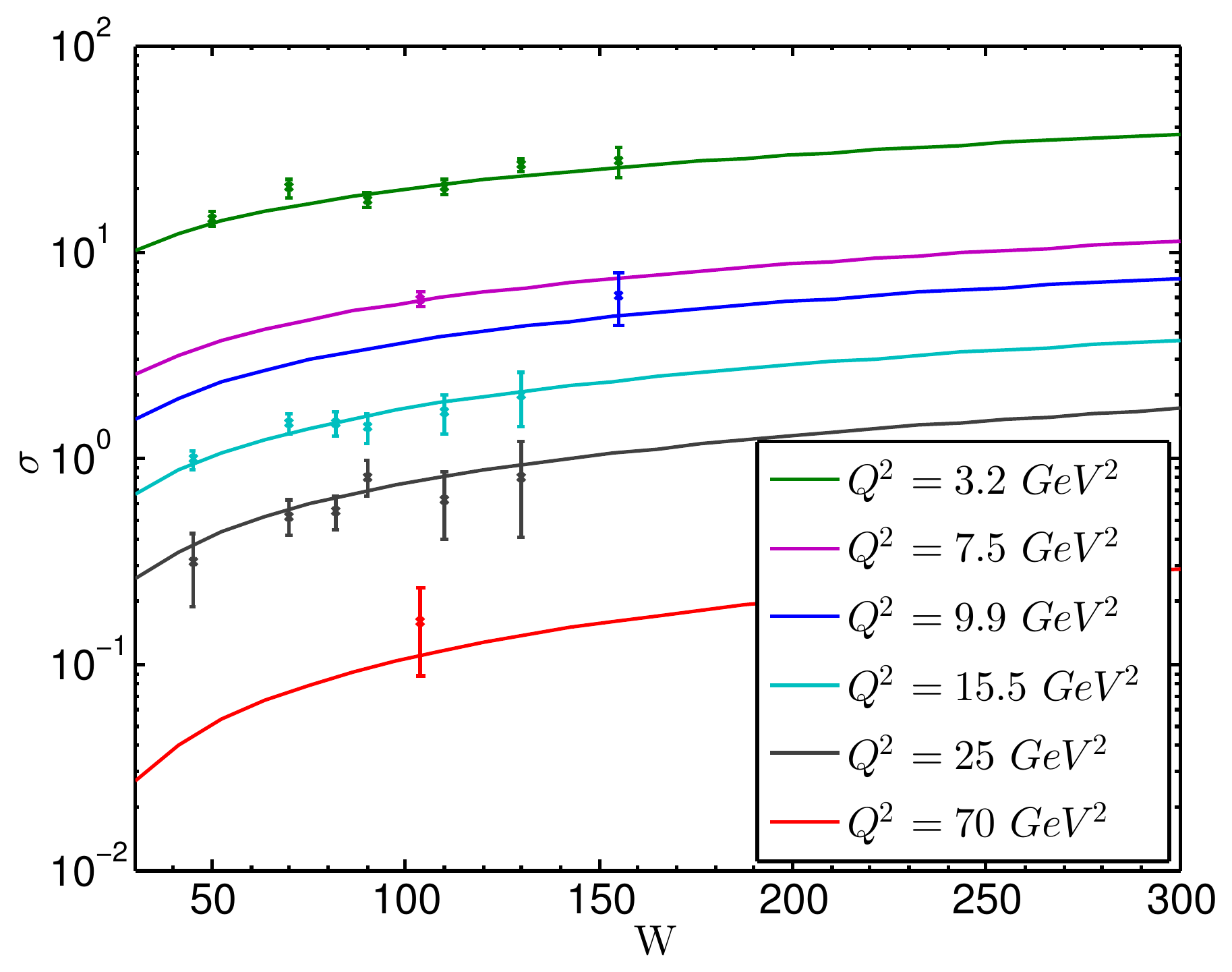}
\caption{{\small The fit of the conformal pomeron model to the data. The first five figures
are for  the differential cross section data and the last one for the cross section.
$W$ is the center of mass energy and we use units of ${\rm GeV}$.
To avoid cluttering the last figure we did not plot all of the $Q^2$  values.}}
\label{fig:conf_dsdt}
\end{center}
\end{figure}

The parameters we fit for the conformal pomeron are the coupling to external states $g_0^2$ (in fact $Cg_0^2$)\footnote{Note that the definition of $g_0$ is different than that in \cite{Brower:2010wf}, so the values for this parameter are not directly comparable, but they can be simply related to each other.}, the radial position of the proton 
$z_*$ and $\rho$, which is related to the intercept by $j_0 = 2 - \rho$. 
For the differential cross section the values we get are
\begin{equation}
\label{eq:dsdt_conf_pars}
g_0^2 = 1.95 \pm 0.85\,,\ \ \ \ z_* = 3.12 \pm 0.160\, {\rm GeV}^{-1}\,,\ \ \ \ \rho = 0.667 \pm 0.048\,.
\end{equation}
corresponding to a $\chi^2$ of
\begin{equation}
  \label{eq:dsdt_conf_chsq}
  \chi^2_{d.o.f.} = 1.33\, .
\end{equation}
We can see in figure \ref{fig:conf_dsdt} the comparison of our model to data. Even though $-t$ is larger than $\Lambda_{QCD}^2$ for all values where we have data, at the smallest values it is close, and there might be some lingering effects of confinement felt. If we exclude the lowest value of $|t|$ from each graph, we get a better fit, with a $\chi^2$ of
\begin{equation}
  \chi^2_{d.o.f.} = 0.76\, .
\end{equation}

For the cross section, we fit the same three parameters. The values we get are
\begin{equation}
  g_0^2 = 8.79 \pm 4.17\,,\ \ \ \ z_* = 6.43 \pm 2.67\ {\rm GeV}^{-1}\,,\ \ \ \ \rho = 0.816 \pm 0.038\, .
\label{eq:pars_sigma_conf}
\end{equation}
Here the integration over $t$ is likely to reduce the weight of the region of low $t$ to the data, so that 
confinement effects are less important, resulting in a better fit  with a $\chi^2$ of
\begin{equation}
  \chi^2_{d.o.f.} = 1.00\, .
\end{equation}
%
%
\noindent In figure \ref{fig:conf_dsdt} we see how the model compares with data. We see that we have a very good fit. Note that in the figure we skip over some values of $Q^2$ in order to avoid cluttering the graph.

\subsection{Eikonalization}\label{sub_sec:data:eikonal}

As mentioned in section \ref{sub_sec:gamma:eikonal} fitting the eikonal model to data does not improve our fits. 
Recalling that $g_0^2$ in (\ref{eq:dsdt_conf_pars}) is actually $Cg_0^2$, with $C$ given by the reference value  (\ref{C}), 
the value for the parameters we obtain for the eikonal fit does not change much from that in (\ref{eq:dsdt_conf_pars}). The only
difference is that the eikonal fit gives a $g_0^2 \lesssim 0.05$ and a value of $C$ that compensates for that change, keeping $C g_0^2$
approximately constant. Indeed, by running the fit for the eikonal model the values of $\rho$ and $z_*$ are the same as those in equations (\ref{eq:dsdt_conf_pars}) and (\ref{eq:pars_sigma_conf}). Now it turns out that once $g_0^2 \approx 0.5$ or lower, the eikonal fit does not change much because we enter the regime where 
the phase shift in the exponential is small, and the amplitude dominated by single pomeron exchange. It is nevertheless  
encouraging that the estimated value for $C$ has the order of magnitude of (\ref{C}). The very same order of magnitude for the constant $C$ was also obtained 
in the black disk fitting (\ref{eq:dsdt_bd_pars}). In both cases there are large uncertainties in the estimated value of $C$.

\begin{figure}[htbp]
\begin{center}
\includegraphics[scale=0.43]{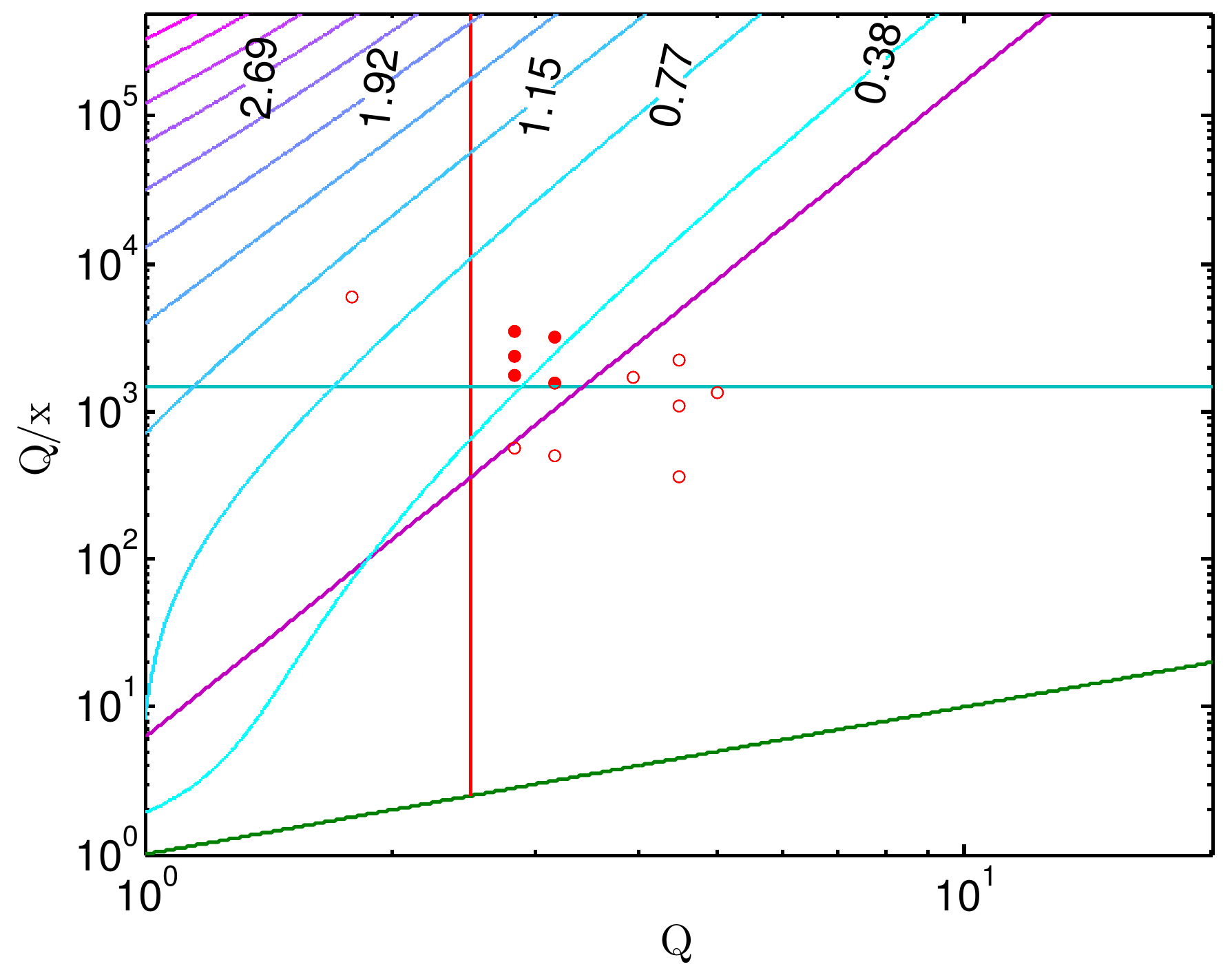}
\caption{{\small Contour plot of the phase shift $\chi$. $Q$ in ${\rm GeV}$.}}
\label{fig:eik_bd}
\end{center}
\end{figure}

It is also useful to analyse the size of $\chi$ for single pomeron exchange in the kinematical region of validity of the   black disk model, figure \ref{fig:eik_bd}.
Due to the aforementioned uncertainty in $C$, it is not the exact values for $\chi$ that are important, but rather the pattern that emerges. We see that the lines of equal $\chi$ are roughly parallel to the line $\log(Q z_*)$, as expected, and the black disk fits in the range where $\chi$ grows, excluding the points with a small value of $\chi$. Also excluded is the region where we believe confinement will start to play a role. 

The same conclusions are supported from the analysis of the total cross section.

\subsection{Hard wall pomeron}\label{HWP}

\begin{figure}[htbp]
\begin{center}
\includegraphics[scale=0.43]{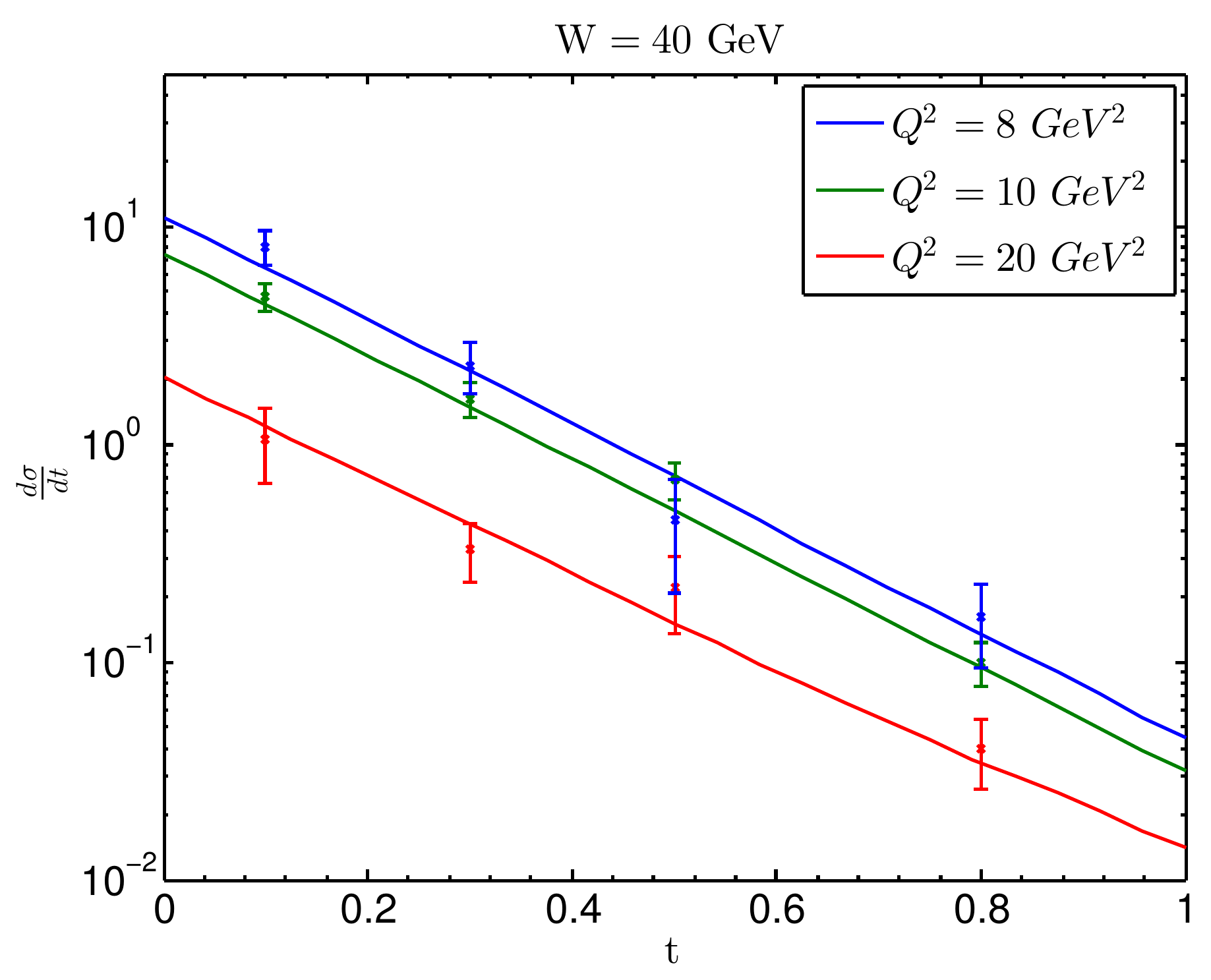}
\includegraphics[scale=0.43]{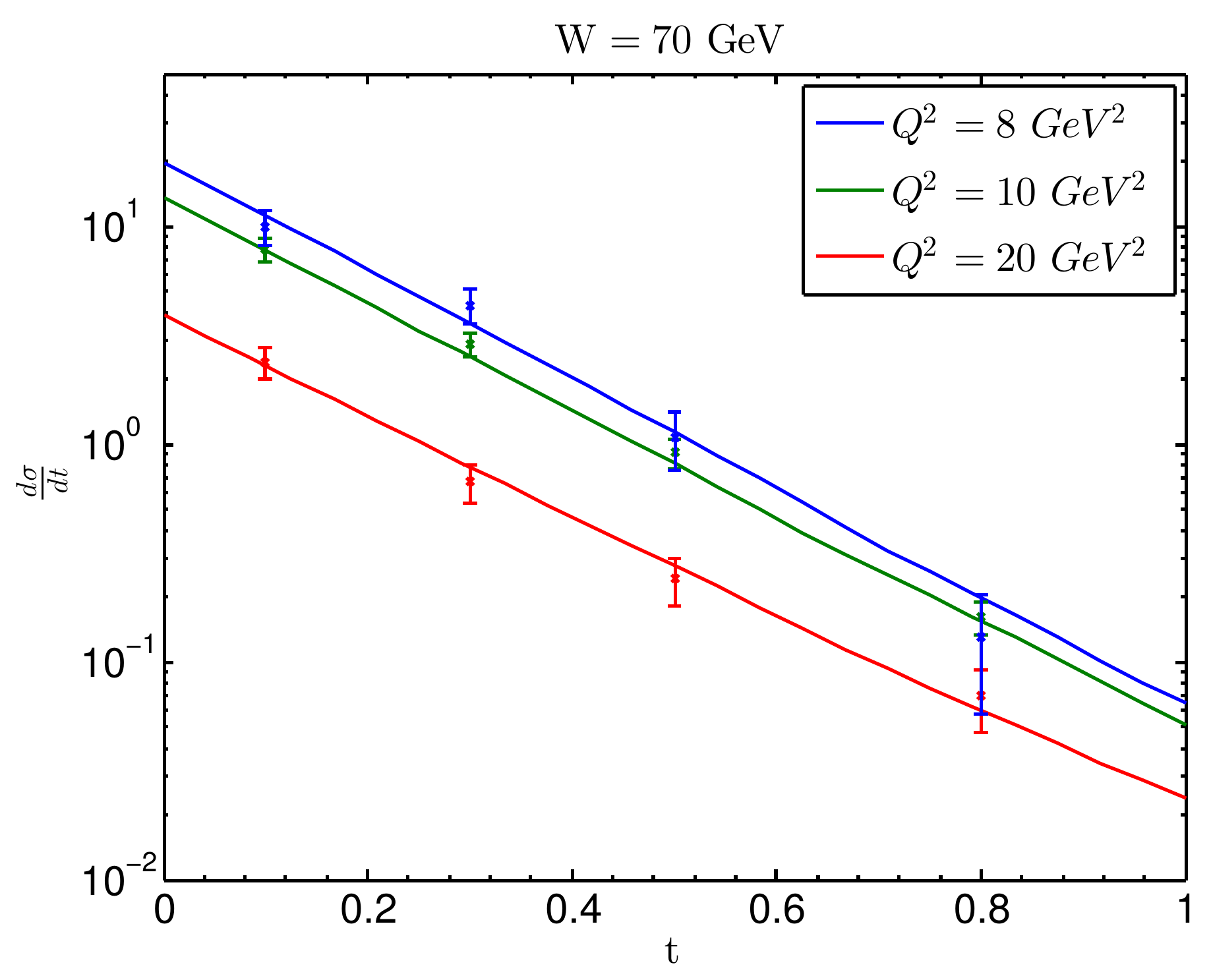}\\
\includegraphics[scale=0.43]{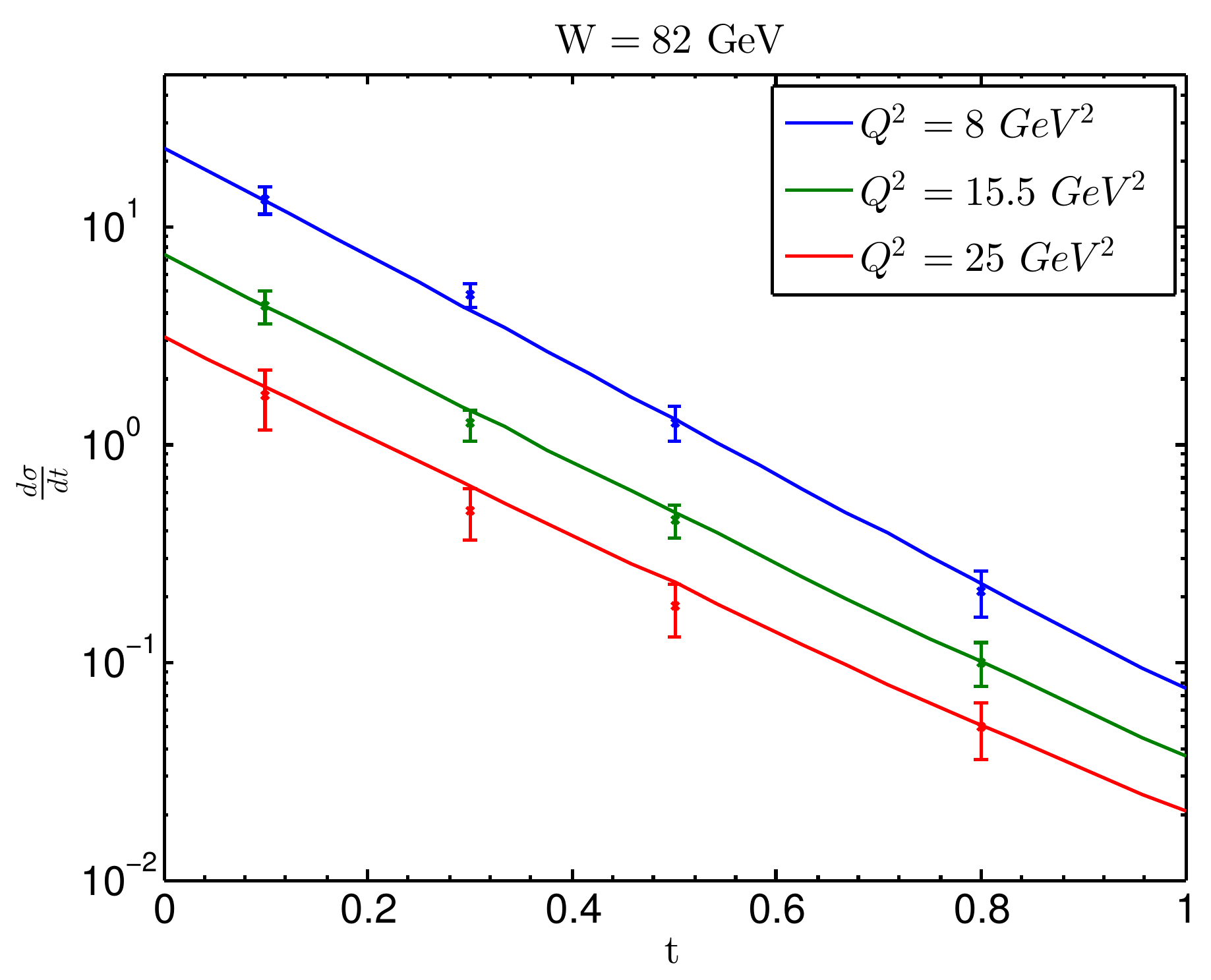}
\includegraphics[scale=0.43]{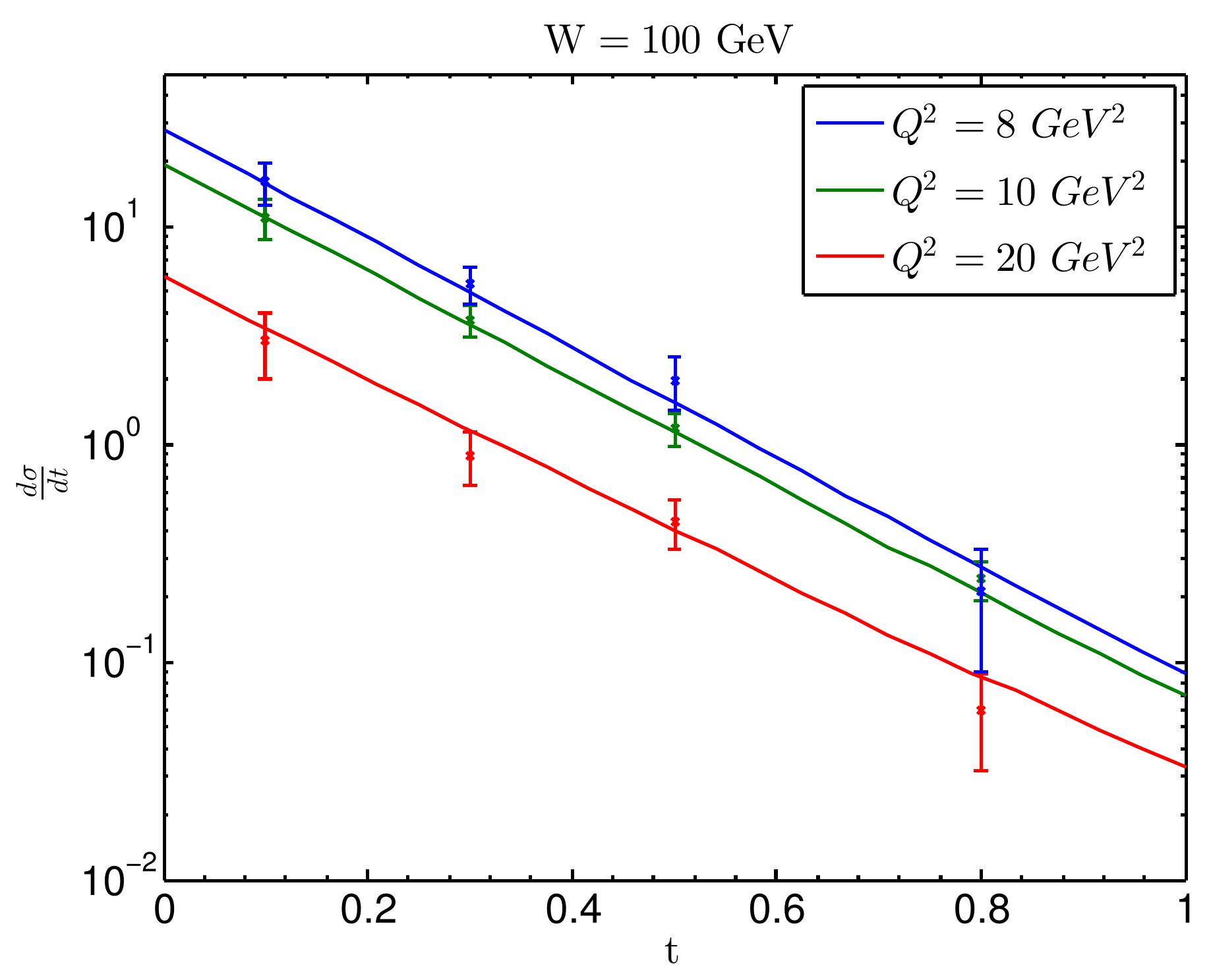}\\
\includegraphics[scale=0.43]{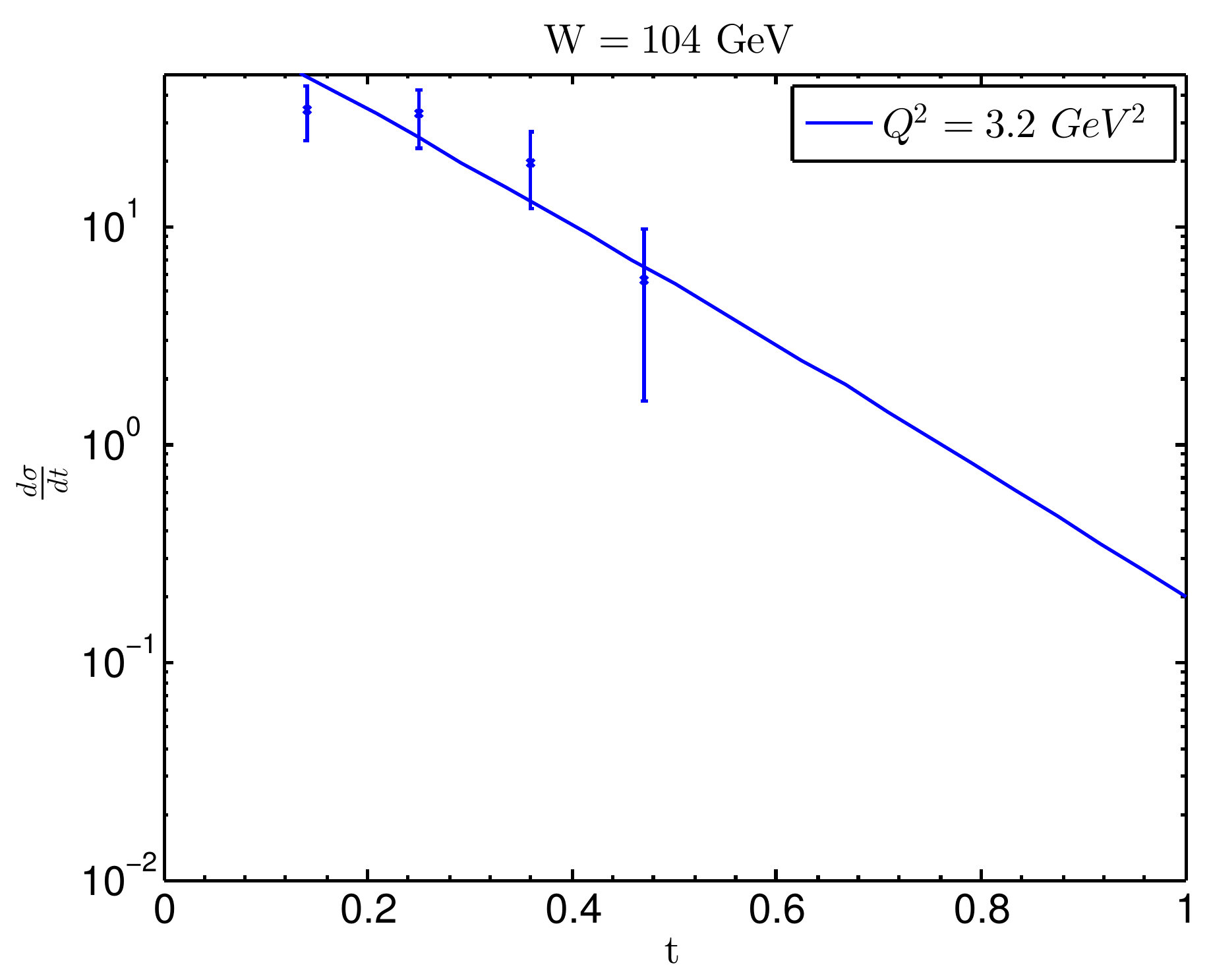}
\includegraphics[scale=0.43]{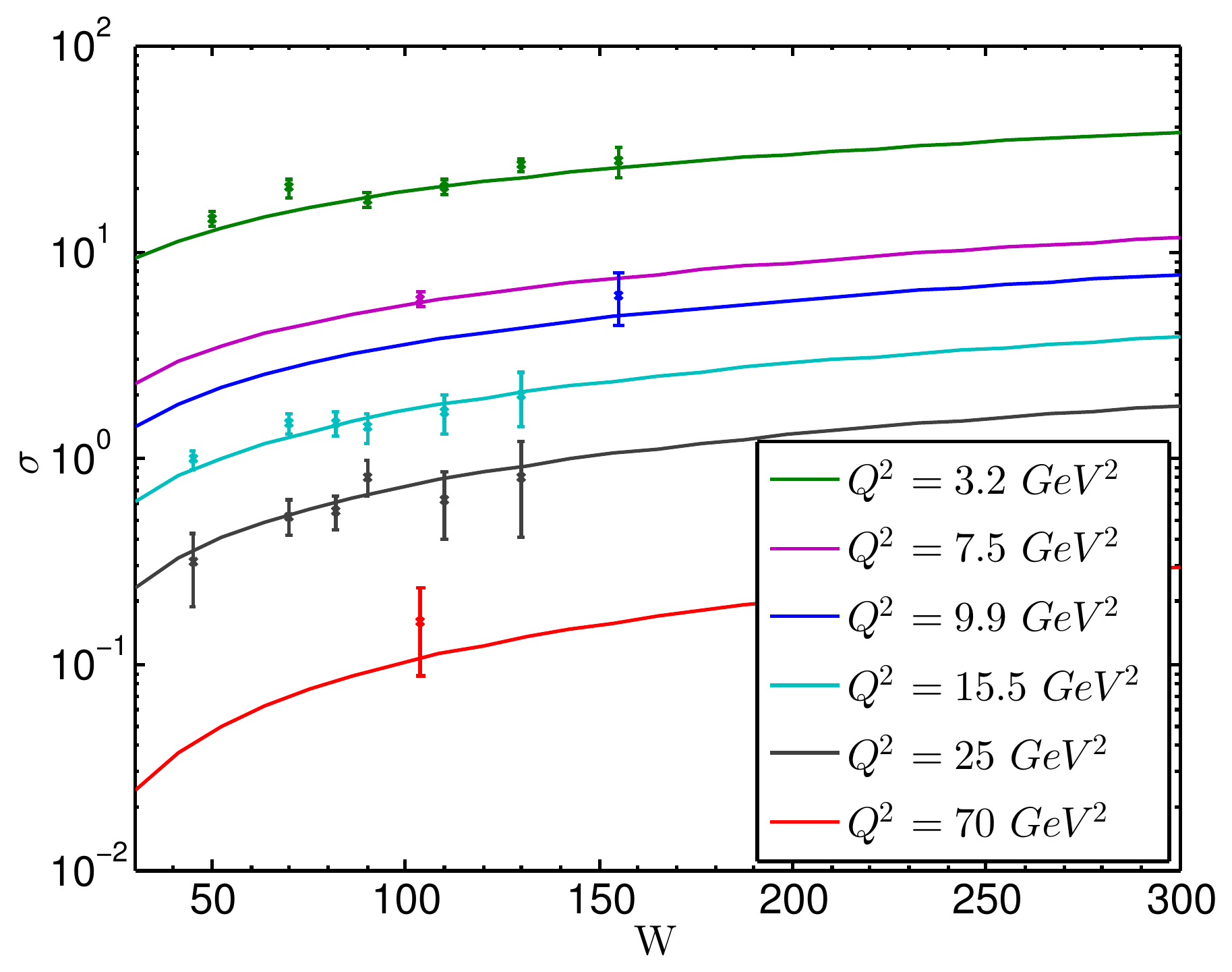}
\caption{{\small 
The fit of the hard-wall pomeron model to the data. The first five figures
are for  the differential cross section data and the last one for the cross section.
$W$ is the center of mass energy and we use units of ${\rm GeV}$.
To avoid cluttering the last figure we did not plot all of the $Q^2$  values.}}
\label{fig:hw_dsdt}
\end{center}
\end{figure}

As mentioned earlier, the hard wall pomeron adds an additional parameter, the position $z_0$ of the cut off in AdS. The parameters we obtain by fitting are
\begin{equation}
  \label{eq:dsdt_hw_pars}
g_0^2 = 2.46\, \pm  0.70\,,\ \ \ z_* = 3.35 \pm 0.41\ {\rm GeV}^{-1},\ \ \ \rho = 0.712 \pm 0.038\,,\ \ \ z_0 = 4.44 \pm 0.82\ {\rm GeV}^{-1}.
\end{equation}
corresponding to a $\chi^2$ of 
\begin{equation}
  \label{eq:dsdt_hw_chsq}
  \chi^2_{d.o.f.} = 0.51\, .
\end{equation}
The fit is clearly better than the conformal one, as can also be seen from figure \ref{fig:hw_dsdt}. The main improvement that it offers is taking into account the effects of confinement, which are still felt at low $|t|$. 
For the cross section (also in figure \ref{fig:hw_dsdt}), the values for the parameters we get are
\begin{equation}
  \label{eq:sigma_hw_pars}
g_0^2 = 6.65 \pm 2.30\,,\ \ \ z_* = 4.86 \pm 2.87\ {\rm GeV}^{-1} ,\ \ \ \rho = 0.811 \pm 0.036\,,\ \ \  z_0 = 8.14 \pm 2.96\ {\rm GeV}^{-1}.
\end{equation}
corresponding to a $\chi^2$ of
\begin{equation}
  \chi^2_{d.o.f.} = 1.03\, .
\end{equation}

Based on the fit to DIS data \cite{Brower:2010wf}, we expect the effects of confinement only to become important once $Q^2$ is about $1\ {\rm GeV}^2$ or lower, and above those energies the conformal and hard-wall models to give very similar results. The above fits confirm these expectations, both with the $\chi^2$ values and the fact that the fitted paramaters for both models are close to each other. 

\section{Conclusion}
\label{Conc}

One of the major goals of particle physics is to accurately calculate and predict cross sections for scattering experiments. Therefore novel techniques that allow us to do such calculations are always welcome, both from a practical point of view, and a theoretical one, since they deepen our understanding of particle interactions. Gauge/string duality already has many interesting applications, and in this paper we have carried out further steps in using it to study diffractive processes in QCD.

Indeed,  pomeron exchange is the dominant contribution to the cross section in low $x$ physics. When all scales are above $\Lambda_{QCD}$, we can explore conformal
Regge theory and compute the cross section arising from one reggeon exchange. This leads to the conformal impact parameter representation (\ref{IPrep}) 
for the transition amplitude $T_{ab}(k_i)$, 
where the conformal amplitude ${\cal B}(S,L)$ can be computed both at weak coupling (\ref{AmplitudeFinalWeak}) and at strong coupling (\ref{AmplitudeFinal}). 
The approximate conformal symmetry in processes dominated by pomeron exchange, when the amplitude grows beyond the single pomeron exchange, 
then leads to the AdS black disk model proposed in \cite{Saturation} and successfully tested against DIS data
inside  the saturation region. 
To extend the analysis to a larger kinematical range one could consider either the weak or the strong coupling expansion for reggeon exchange, and 
then include confinement effects that break conformal invariance. In this paper we considerer the strong coupling expansion for the exchange of a pomeron,
first identified by Brower, Polchinski, Strassler and Tan in  \cite{Brower1} as the graviton Regge trajectory in AdS, or the BPST pomeron. 
Its advantage, as compared to the perturbative BFKL pomeron, is that it unifies the {\em soft} and the {\em hard} pomerons through diffusion in AdS space, which was also shown explicitly in a comparison to data in \cite{Brower:2010wf}. This allows for comparison in a much larger kinematical range, unaccessible to traditional perturbative QCD methods, including for energy scales close to $\Lambda_{QCD}$, where the processes are inherently non-perturbative. The pomeron's universality in a very wide range of high energy scattering processes then makes such a study very useful.

After reviewing and further elucidating some of the properties of the conformal pomeron Regge theory, we computed the hadronic Compton tensor and considered three
models:  the AdS black disk model, the conformal pomeron model and the hard wall pomeron model, which incorporates confinement.
Some of the computations have already been done when comparing to DIS data for the black disk model \cite{Saturation} and both pomeron models \cite{Brower:2010wf},  but we
made several improvements necessary to compute the amplitude when $t \neq 0$, as required for deeply virtual Compton scattering. 
This is the natural next step after DIS, due to the similarity between both processes.

In section \ref{Data} we carried out the data comparison for the above three models.
We found a very good agreement with data, where each of the models is applicable. 
We also found that the parameters are in the same range as those found in the previous works \cite{Saturation,Brower:2010wf}, 
with the parameters for the cross section being the closest. 
This is natural, since DIS also measures the total cross section. Furthermore, all the main previous conclusions are supported:
inside saturation one observes a conformal  (AdS) black disk \cite{Saturation}; 
the conformal pomeron fits the data well in the conformal region, the hard wall model becomes more accurate when
$Q$ and $t$ approach the QCD scale in a region where confinement effects become important before saturation starts to occur \cite{Brower:2010wf}.  
  
Our analysis in sections \ref{sub_sec:data:conf_pomeron} and \ref{HWP} further suggests that confinement effects become important when $-t$ is of order $0.1\ {\rm GeV}^2$, and the hard-wall model can explain this data as well, further cementing its status as a useful tool in studying the string duals of confining theories, such as QCD. 

\begin{figure}[htbp]
\begin{center}
\includegraphics[scale=1]{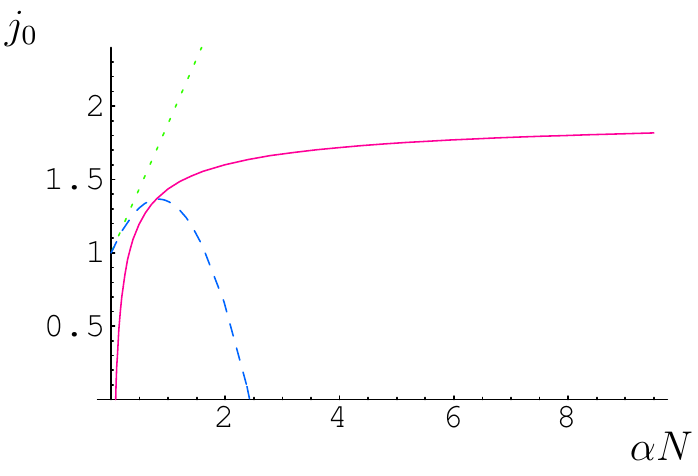}
\caption{{\small The comparison of the pomeron intercept for weak and strong coupling. The green and blue dotted lines give  the dependence 
of the pomeron intercept to first and second order on the 't Hooft coupling $\alpha N$, respectively. 
The solid line gives the strong coupling dependence obtained from strings in  AdS. Figure taken from reference \cite{Brower1}.}}
\label{fig:intercept}
\end{center}
\end{figure}

Furthermore, as for DIS,  the fit for the pomeron intercept again has a value in the crossover region between strong and week coupling, $j_0=1.2-1.3$.  
The value we obtain for the intercept, as a function of the 't Hooft coupling is represented in figure \ref{fig:intercept},
together with the corresponding weak and strong coupling curves. Since we are
in an intermediate region, it is very possible that  the weak coupling expansion is as good as the strong coupling one. This raises the question whether
one can construct a weak coupling hard wall model. Indeed, an attempt to fit DIS data using the BFKL pomeron with a IR cut-off
appeared in \cite{Kowalski:2010ue}, but the strong coupling model fits data better in the IR region. It is therefore possible that 
using the (stringy) AdS dual as a tool to introduce a IR cut-off, just as in the strong coupling case, will give an equally good fit to data.  
We leave this question to future research.

Our goal is to develop a framework for a study of a wide range of high energy scattering processes, and eventually use all these processes to constrain the parameters of our model. For example, other processes where we can apply an analogous study are vector meson production and also proton-proton total cross sections. It is also important to move beyond $2\rightarrow 2$ scattering. Indeed, work on applying pomeron exchange in AdS to double diffractive Higgs production, a $2\rightarrow 3$ process, should appear shortly \cite{Brower:2012mk}. 

We have taken here an important step for such a unified framework, but though much is taken, much abides. As a next step we plan to extend this study to 
vector meson production, where the coupling of the mesons to the pomeron is determined by the IR physics deep inside AdS. This will be an important
study, among other reasons for allowing a better examination of the interplay between confinement and saturation. 
Another useful project is to numerically calculate the hard-wall pomeron propagator in impact parameter space, thus improving the approximation we used here.
It would likewise be quite interesting to use a different model of confinement, allowing us to identify which of its features are model independent.
We expect to return to such questions in the near future.


\bigskip

\begin{center} 
{\bf Acknowledgements} 
\end{center}
The authors are grateful to  Jo\~ao Penedones and Chung-I Tan for helpful discussions. This work was partially funded by grants PTDC/FIS/099293/2008 and CERN/FP/116358/2010.
\emph{Centro de F\'{i}sica do Porto} is partially funded by FCT. The work of M.D. is supported by the FCT/Marie Curie Welcome II program.

\newpage

\appendix

\section{Graviton exchange in AdS \label{SC}}

The impact parameter representation (\ref{IPrep}) assumes conformal symmetry and Regge kinematics. 
In this section we compute this amplitude using the AdS/CFT duality, in particular we check the non-trivial 
form of the tensor function (\ref{Psi}) associated to the coupling of the vector current (dual to a $U(1)$ AdS gauge field) to the pomeron (dual to the graviton
Regge trajectory).
To fix our notation we start by writing the action for the gravity fields
\begin{equation}
S= \frac{1}{2\kappa^2} \int d^5 x \sqrt{-g} \left(  {\cal R} - 12\right) + S_m\,.
\end{equation}
In our units the AdS vacuum has radius 1.
In the context of ${\cal N}=4$ SYM the gravitational coupling $\kappa=2\pi/N$ and the  AdS radius  $R^4= 4\pi g_s N\alpha'^2$.
For the matter fields we shall consider a $U(1)$ gauge field $A$, dual to the electromagnetic current operator, and a complex scalar field
with mass given by $m^2=\Delta(\Delta-4)$, with a bound state dual to the target hadron (after introducing a gauge/gravity model for confinement). 
The matter fields action is therefore given by
\begin{equation}
S_m= \int d^5 x \sqrt{-g} \left( 
 -\frac{1}{4}\,F^2 - \frac{1}{2} (\partial \phi )^2 - \frac{1}{2} \Delta(\Delta-4)\phi^2\right)  \,,
\end{equation}
where  $F=dA$.

\subsection{Propagators}

We denote the bulk to boundary propagator for the scalar field by $\Pi(X,y)$ where $X=(z,x)$ are the Poincar\'e coordinates of the bulk point and $y$ the coordinate of the
boundary point, with $x,y \in \bb{M}^4$.  We have
\begin{equation}
\Pi(X,y) = \frac{\Gamma(\Delta)}{\pi^2 \Gamma(\Delta-2)}\, \left( \frac{z}{z^2+w^2} \right)^\Delta\,,
\end{equation}
with $w=x-y$.  It will be useful below to have the Fourier transform of this propagator 
\begin{equation}
\Pi(z,k) = \int d^4 w\, e^{ik\cdot w} \, \Pi(X,y) =  \frac{2^{3-\Delta}}{\Gamma(\Delta-2)} \,Q^{\Delta-2} z^2K_{2-\Delta}(Qz)\,,
\label{ScalarPropFourier}
\end{equation}
where $Q=\sqrt{k^2}$. Taking the boundary limit we have $z^{\Delta-4}\Pi(X,y) \rightarrow  1\cdot\delta(x-y)$. This normalisation fixes the constant 
$\bar{C}$ in (\ref{2pfNormalization}) to
\begin{equation}
\bar{C}= 2\,\frac{\Delta-2}{\Delta}\,\frac{\Gamma(\Delta+1)}{\pi^2 \Gamma(\Delta-2)}\,.
\label{Cbar}
\end{equation}

For the  bulk to boundary propagator of the $U(1)$ gauge field  $\Pi_{\mu a}(X,y)$, working in the Lorentz gauge $D^\mu \Pi_{\mu a}=0$, we have
\begin{equation}
\Pi_{b a}(X,y) = \frac{1}{2\pi^2}\, \frac{ \left(3z^2 -w^2\right) \eta_{ab} + 4w_aw_b }{(z^2 +w^2)^3}  \,,
\ \ \ \ \ \ \ \ \ \ \
\Pi_{z a}(X,y) = 0\,,
\end{equation}
where again $w=x-y$. The Fourier transform is then
\begin{equation}
\Pi_{b a}(z,k) = \int d^4 w\, e^{ik\cdot w} \, \Pi_{b a}(X,y) = \left( \eta_{ab} - \frac{k_a k_b}{k^2}\right) (Qz)K_1(Qz)\,.
\label{GaugePropFourier}
\end{equation}
It is now trivial to see that taking the limit near the boundary the propagator generalises 
$\Pi_{b a}(X,y) \rightarrow  \eta_{ab}\cdot\delta(x-y)$, such that the Lorentz gauge condition $\partial^b \Pi_{b a}=0$ is satisfied
on the boundary. Computing the two point function we obtain a normalisation in  (\ref{2pfNormalization}) with
\begin{equation}
C=\frac{6}{\pi^2}\,.
\end{equation}

Let us finally consider the bulk to bulk propagator for the graviton. In the Regge limit the propagator equation for a spin $2$ and dimension $4$ field
simplifies to (see \cite{Cornalba:2007zb,Penedones:2007ns} for details)
\begin{equation}
\left(  \Box +2  \right) \Pi_{\alpha\beta\mu\nu} (X,\bar{X})= i g_{\alpha\left(\mu\right.} g_{\left.\nu\right)\beta} \, \delta_{AdS}(X,\bar{X}) + \cdots\,.
\end{equation}
where $\cdots$ represent terms that are not important in the Regge limit. As usual, given the energy-momentum source $T^{\mu\nu}$, the metric perturbation is given by
\begin{equation}
h_{\alpha\beta}(X) = i \kappa \int d^5 \bar{X} \sqrt{-g} \,\Pi_{\alpha\beta\mu\nu} (X,\bar{X}) \, T^{\mu\nu}(\bar{X}) \,.
\label{GravitonCoupling}
\end{equation}
In the following discussion we will not need the full form of the graviton propagator. Instead we will only need to know the integrated
propagator 
\begin{equation}
\int d\bar{x}^+ dx^-  \Pi_{++--}  (X,\bar{X})  = -2i (z \bar{z})^{-1} \,\Pi_\perp(L) \,,
\label{H3Prop}
\end{equation}
where we use light cone coordinates $x=(x^+,x^-,x_\perp)$ on the $\bb{M}^4$ boundary and similarly for $\bar{x}$.
$\Pi_\perp(L)$ is the scalar propagator in $H_3$ with mass squared 3,
and it depends only on the invariant $L$ given in (\ref{S,L}) with $l_\perp = x_\perp - \bar{x}_\perp$. In other words, $\Pi_\perp$ satisfies the Euclidean
equation 
\begin{equation}
\big(  \Box_{H_3} - 3 \big) \,\Pi_\perp(x,\bar{x}) = - \delta_{H_3}(x,\bar{x})\,,
\end{equation}
where $x=(z,x_\perp)$ and $\bar{x}=(\bar{z},\bar{x}_\perp)$ in Poincar\'e coordinates on $H_3$. Explicitly we have
\begin{equation}
\Pi_\perp(L) = \frac{1}{2\pi}\,\frac{e^{-3L}}{1-e^{-2L}}\,.
\end{equation}

\subsection{Witten diagram}

We have now the necessary ingredients to evaluate the Witten diagram that computes the correlation function (\ref{4pf}) in the Regge limit, which is dominated by 
the t-channel graviton exchange. Feynman rules in AdS give for  $G_{ab}(y_i)=\left\langle j_a(y_1) {\cal O}(y_2) j_b(y_3) {\cal O}(y_4) \right\rangle $
the expression
\begin{align}
A_{ab}(y_i)  = 
(i\kappa)^2 \int & \frac{d^5X}{z^5}\,\frac{d^5\bar{X}}{\bar{z}^5}\, 
2\partial_{\left[ -\right.}\Pi_{\left.\alpha\right] a}(X,y_1)\, g^{\alpha\beta}(z)\, 2\partial_{\left[ -\right.}\Pi_{\left.\beta\right] b}(X,y_3)
\nonumber
\\&
(16z ^4\bar{z} ^4)\, \Pi_{++--}(X,\bar{X})\, \partial_+\Pi(\bar{X},y_2)\,\partial_+\Pi(\bar{X},y_4)\,,
\label{WittenDiagram}
\end{align}
where we used that the graviton couples to the energy momentum tensor $T_{\mu\nu}$ as given in (\ref{GravitonCoupling}), and we note that the
factor $16z^4\bar{z}^4$ appears from metric elements $g^{+-}$ computed at $z$ and $\bar{z}$.
Evaluating the Fourier transform
\beq
A_{ab}(k_i)   = \int \left(  \prod_{i=1}^4 d^4y_i e^{-i k_i\cdot y_i} \right)  A_{ab}(y_i)   \,,
\eeq
we obtain
\begin{align}
A_{ab}(k_i)   = -\kappa^2 \int &  \frac{d^5X}{z^5}\,\frac{d^5\bar{X}}{\bar{z}^5}\, 
e^{-i(k_1+k_3) \cdot x}\,
2 k_{1\left[-\right.}
\Pi_{\left.\alpha\right] a}(z,k_1)\, g^{\alpha\beta}(z)\, 2k_{3\left[-\right.}\Pi_{\left.\beta\right] b}(z,k_3)
\nonumber
\\&
(16z^4\bar{z}^4)\,  \Pi_{++--}(X,\bar{X})\, e^{-i(k_2+k_4) \cdot \bar{x}}  \,
k_{2+}\Pi(\bar{z},k_2)\, k_{4+}\Pi(\bar{z},k_4)\,,
\end{align}
where the momentum space bulk to boundary propagators are given in (\ref{ScalarPropFourier}) and (\ref{GaugePropFourier}).
Next we insert the external kinematics, as given in (\ref{k1k2}) and (\ref{k3k4}). In the Regge limit the result simplifies to
\begin{align}
A_{ab}(k_i)   = &- \kappa^2 (2\pi)^4\, \delta(0)
\int \,dl_\perp\, e^{i q_\perp \cdot l_\perp }
\int   \frac{dz}{z^3}\,\frac{d\bar{z}}{\bar{z}^3} \,
2 k_{1\left[-\right.}\Pi_{\left.\alpha\right] a}(z,k_1)\,g^{\alpha\beta}(z)\, 2k_{3\left[-\right.}\Pi_{\left.\beta\right] b}(z,k_3)
\nonumber
\\&
k_{2+}\Pi(\bar{z},k_2)\, k_{4+}\Pi(\bar{z},k_4)\left(  \int\frac{d\bar{x}^+dx^-}{2}  \left(16z^2\bar{z}^2\right) \Pi_{++--}(X,\bar{X}) \right)
\,,
\end{align}
where $l_\perp=x_\perp-\bar{x}_\perp$\,. We replaced the space-time volume by the momentum conserving $\delta$ function,
$V= \left(2\pi^4\right) \delta(0)$, to extract the amplitude $T_{ab}$ introduced in (\ref{4pf}). 
The integral inside the large parenthesis gives precisely the scalar propagator in $H_3$ given in  (\ref{H3Prop}).
Thus, we obtain finally
\begin{equation}
T_{ab}(k_i)   =  - \kappa^2 \int \,dl_\perp\, e^{i q_\perp \cdot l_\perp }
\int   \frac{dz}{z^3}\,\frac{d\bar{z}}{\bar{z}^3} \,
z^4 \left( t_{ab} \Psi_0 + u_{ab}  \Psi_1    \right) 
\frac{s}{4}\,
 \Pi(\bar{z},k_2)\,\Pi(\bar{z},k_4)    ( 16   z\bar{z}) \,\Pi_\perp(L)\,,
\end{equation}
where  $\Psi_n$ is defined in (\ref{Psi_n}) and
\begin{align}
t_{ab}  &= \left( Q^2 \eta_{-a}  -  k_{1-} k_{1a} \right) \left( Q'^2 \eta_{-b}  -  k_{3-} k_{3b} \right)  \,,
\nonumber
\\
u_{ab}&= QQ' \big(k_{1-} k_{3-} \eta_{ab} - k_{1-} k_{3a} \eta_{-b} -k_{1b} k_{3-} \eta_{-a} +\left(k_1\cdot k_3\right) \eta_{-a} \eta_{-b}\big) \,.
\end{align}
A simple computation shows that
\beq
t^{ab}
= 
\frac{s}{4}\left(
\begin{array}{ccc}
s&Q'^2 + q_\perp^2& \sqrt{s} q_\perp^i\\ \\
Q^2&\frac{1}{s} \, Q^2 \left( Q'^2 + q_\perp^2 \right)& \frac{1}{\sqrt{s}} \, Q^2 q_\perp^i\\ \\
0&0&0
 \end{array}
\right)\,.
\ \ \ \ \ \ \
u^{ab}
= 
\frac{s}{4}\left(
\begin{array}{ccc}
0&0&0\\ \\
0&0&0\\ \\
0& \frac{2}{\sqrt{s}}\,q_{\perp}^i & \delta_{ij}
 \end{array}
\right)\,.
\label{tu}
\eeq
 Thus we recover the tensor function defined in (\ref{Psi}),
 \beq
 \Psi^{ab}(z) = \frac{4 z^4}{s} \left( t^{ab} \Psi_0 + u^{ab}  \Psi_1    \right) \,,
 \eeq
 with $C=6/\pi^2$. Also, using $\bar{C}$ as given in (\ref{Cbar}) we recover the scalar function in (\ref{Phi}), more precisely 
\beq
\Phi(\bar{z}) =\Pi(\bar{z},k_2)\,\Pi(\bar{z},k_4)\,.
 \eeq
Finally we can read the strong coupling phase shift associated to graviton exchange
\beq
\chi(S,L) =   - \frac{\kappa^2}{2} S \,\Pi_\perp(L)\,,
\eeq
in agreement with \cite{Cornalba:2007zb,Brower:2007qh}. This computation is a non-trivial check of the impact parameter
representation introduced in \cite{CCP06}, which only relies on conformal symmetry.

\end{document}